\documentclass[11pt,a4paper]{article}
\usepackage{sint,amsfonts}
\usepackage{lineno,hyperref}
\usepackage{booktabs}
\usepackage{amsmath}
\usepackage{amssymb}
\usepackage{bbold}
\usepackage{wasysym}
\usepackage[flushleft]{threeparttable}
\usepackage{lscape}
\usepackage{rotating}
\usepackage{simplewick}
\usepackage{upgreek}
\usepackage[english]{babel}
\usepackage{cite}

\def\rms{{\rm s}}
\def\rmc{{\rm c}}

\def\cm{\rmc_\mu}
\def\cn{\rmc_\nu}
\def\ca{\rmc_\alpha}
\def\crh{\rmc_\rho}
\def\cs{\rmc_\sigma}
\def\cl{\rmc_\lambda}
\def\sm{\rms_\mu}

\def\sl{\rms_\lambda}

\def\bsx{{\boldsymbol x}}
\def\bsxi{{\boldsymbol \xi}}

\def\proof{\noindent{\sl Proof:}\kern0.6em}

\def\frac#1#2{\hbox{$#1\over#2$}}
\def\dual{\mathstrut^*\kern-0.1em}

\def\lvec#1{\setbox0=\hbox{$#1$}
    \setbox1=\hbox{$\scriptstyle\leftarrow$}
    #1\kern-\wd0\smash{
    \raise\ht0\hbox{$\raise1pt\hbox{$\scriptstyle\leftarrow$}$}}
    \kern-\wd1\kern\wd0}
\def\rvec#1{\setbox0=\hbox{$#1$}
    \setbox1=\hbox{$\scriptstyle\rightarrow$}
    #1\kern-\wd0\smash{
    \raise\ht0\hbox{$\raise1pt\hbox{$\scriptstyle\rightarrow$}$}}
    \kern-\wd1\kern\wd0}
\def\slash#1{\setbox0=\hbox{$#1$}\setbox1=\hbox{$\kern1pt/$}
    #1\kern-\wd0\kern1pt/\kern-\wd1\kern\wd0}

\def\nab#1{{\nabla_{#1}}}
\def\nabstar#1{{\nabla\kern0.5pt\smash{\raise 4.5pt\hbox{$\ast$}}
               \kern-5.5pt_{#1}}}

\def\nabbar#1{{\overleftarrow{\nabla}_{#1}}}
\def\nabbarstar#1{{\overleftarrow{\nabla}\kern0.5pt\smash{\raise 4.5pt\hbox{$\ast$}}
               \kern-5.5pt_{#1}}}

\def\nabdbar#1{{\overleftrightarrow{\nabla}_{#1}}}
\def\nabdbarstar#1{{\overleftrightarrow{\nabla}\kern0.5pt\smash{\raise 4.5pt\hbox{$\ast$}}
               \kern-5.5pt_{#1}}}

\def\drv#1{{\partial_{#1}}}
\def\drvstar#1{{\partial\kern0.5pt\smash{\raise 4.5pt\hbox{$\ast$}}
               \kern-6.0pt_{#1}}}

\def\ldrvstar#1{{\lvec{\,\partial}\kern-0.5pt\smash{\raise 4.5pt\hbox{$\ast$}}
               \kern-5.0pt_{#1}}}

\def\MSbar{\overline{\rm MS\kern-0.5pt}\kern0.5pt}
\def\Nf{{N_f}}

\def\psibar{\overline{\psi}}

\def\zetabar{\bar{\zeta}}
\def\zetaprime{\zeta\kern1pt'}
\def\zetabarprime{\zetabar\kern1pt'}

\def\dirac#1{\gamma_{#1}}
\def\diracstar#1#2{
    \setbox0=\hbox{$\gamma$}\setbox1=\hbox{$\gamma_{#1}$}
    \gamma_{#1}\kern-\wd1\kern\wd0
    \smash{\raise4.5pt\hbox{$\scriptstyle#2$}}}

\def\SUthree{{\rm SU(3)}}
\def\SUn{{\rm SU}(N_c)}

\def\tr{{\rm tr}}
\def\Tr{{\rm Tr}}

\def\Sg{S^{G}}
\def\Sf{S^{F}}

\def\Ds{D_{\rm s}}
\def\DsdagDs{\Ds{\Ds}^{\kern-1pt\dagger}}

\def\avg#1{{\kern1.0pt\overline{\kern-1.0pt#1\kern-1.0pt}\kern1.0pt}}

\newcommand{\be}{\begin{equation}}
\newcommand{\ee}{\end{equation}}
\newcommand{\bea}{\begin{eqnarray}}
\newcommand{\eea}{\end{eqnarray}}

\newcommand{\msbar}{{\rm \overline{MS\kern-0.05em}\kern0.05em}}

\newcommand{\ba}{\begin{eqnarray}}
\newcommand{\ea}{\end{eqnarray}}

\renewcommand{\vec}[1]{\boldsymbol{#1}}

\begin{document}

\begin{titlepage}
\begin{center}
$\;\;\;$
\vspace{1.5cm}

{\Large\bf Non-perturbative definition of the QCD energy-momentum tensor
on the lattice\\[0.5ex]}

\end{center}
\vskip 0.75 cm
\begin{center}
{\large Mattia Dalla Brida$^{\scriptscriptstyle a,b}$, Leonardo Giusti$^{\scriptscriptstyle a,b}$, and
Michele Pepe$^{\scriptscriptstyle b}$}
\vskip 1.25cm
$^{\scriptstyle a}$ Dipartimento di Fisica, Universit\`a di Milano-Bicocca,\\
                 Piazza della Scienza 3, I-20126 Milano, Italy\\ 
\vskip 0.25cm
$^{\scriptstyle b}$ INFN, Sezione di Milano-Bicocca,\\
                 Piazza della Scienza 3, I-20126 Milano, Italy\\

\vskip 1.5cm
{\bf Abstract}
\vskip 0.35ex
\end{center}

\noindent
We present a strategy to define non-perturbatively the energy-momentum tensor in
Quantum Chromodynamics (QCD) which satisfies the appropriate Ward identities and
has the right trace anomaly. The tensor is defined by regularizing the theory on a
lattice, and by fixing its renormalization constants non-perturbatively by suitable
Ward identities
associated to the Poincar\'e invariance of the continuum theory. The latter are derived
in thermal QCD with a non-zero imaginary chemical potential formulated in a moving
reference frame. A renormalization group analysis leads to simple renormalization-group-invariant
definitions of the gluonic and fermionic contributions to either the singlet or the non-singlet components
of the tensor, and therefore of their form factors among physical states.
The lattice discussion focuses on the Wilson discretization of quark fields but the strategy is
general. Specific to that case, we also carry out the analysis for the on-shell
O($a$)-improvement of the energy-momentum tensor. The renormalization and improvement programs
profit from the fact that, as shown here, the thermal theory enjoys de-facto automatic O($a$)-improvement
at finite temperature. The validity of the proposal is scrutinized analytically by a study to 1-loop order
in lattice perturbation theory with shifted and twisted (for quarks only) boundary conditions. The latter
provides also additional useful insight for a precise non-perturbative calculation of the renormalization
constants. The strategy proposed here is accessible to Monte Carlo computations,
and in this sense it provides a practical way to define non-perturbatively the energy-momentum
tensor in QCD. 


\vfill

\eject

\end{titlepage}

\section{Introduction}
The energy-momentum tensor, $T_{\mu\nu}$, is a central quantity in a quantum field 
theory since it groups together the currents associated to the invariance of the
theory under space-time translations, from which also those associated to the
larger Poincar\'e group and scale invariance can be built. Apart the
theoretical one, the great interest in the energy-momentum tensor is manifold.
For instance, in general relativity it enters
the Einstein field equations acting as the source of space-time curvature generated
by the fields. In thermal field theories its expectation values provide the equation
of state of the theory, while its two-point correlation functions allow one
to measure the transport coefficients of the plasma.

The only known non-perturbative regularization of QCD is the lattice where, however,
the Poincar\'e group is explicitly broken into discrete subgroups, and the full
symmetry is recovered in the continuum limit. As a consequence, a given definition of the
energy-momentum tensor on the lattice needs to be properly renormalized to guarantee that
the associated charges are the generators of the Poincar\'e group in the continuum limit,
and that the trace anomaly is correctly reproduced.

In order to construct the renormalized energy-momentum tensor, the way to proceed is to impose
suitable Ward Identities (WIs) at fixed lattice spacing that hold up to cutoff effects which vanish in the continuum 
limit. Indeed that problem was addressed in Refs.~\cite{Caracciolo:1989pt,Caracciolo:1989bu}
for the Yang-Mills theory and QCD, where it was shown that on the lattice the 10-dimensional symmetric
tensor $T_{\mu\nu}$ breaks into the sum of a sextet, a triplet and a singlet representation of the hypercubic group.
Each one
of those three parts picks up finite renormalization constants which were computed to 1-loop order in
perturbation theory~\cite{Caracciolo:1991vc,Caracciolo:1991cp,Burgio:1996ji}, see
also \cite{Capitani:1994qn,Capitani:2002mp,Yang:2016xsb}.
However, it was not clear how to define the renormalization constants so that they could be computed
non-perturbatively.

An important step forward was made a few years ago by noticing that useful
WIs to fix the renormalization constants are obtained by considering the theory in a finite box,
where the Euclidean Lorentz symmetry is also softly broken by its
shape~\cite{Giusti:2010bb,Giusti:2011kt,Giusti:2012yj}.
In particular, if the length in one (temporal) direction $L_0$ is chosen to be finite
(thermal theory), interesting WIs follows. They become particularly simple and of practical
use if the periodicity axes are tilted with respect to the lattice grid (moving reference frame), i.e. if the hard
breaking of the Poincar\'e symmetry due to the lattice discretization and the soft one due to the finite temporal
direction are not aligned. This set-up has a natural implementation in the Euclidean path-integral formulation in
terms of shifted boundary conditions~\cite{DellaMorte:2010yp,Giusti:2010bb,Giusti:2011kt,Giusti:2012yj}.
These ideas led for the first time to a non-perturbative definition of $T_{\mu\nu}$ in the $\SUthree$
Yang-Mills theory~\cite{Giusti:2015daa}.

The purpose of this paper is to present the WIs for defining non-perturbatively the energy-momentum
tensor in QCD. By generalizing  what has been done in the $\SUthree$ Yang-Mills theory, 
we work in the framework of shifted boundary conditions supplemented by the presence
of an imaginary chemical potential. It is the latter that gives us the handle to solve the problem of the mixing
between the gluonic and fermionic parts of the tensor since the chemical potential couples differently
to quark and gluons. Most interestingly, the derived relations can be used in practice to carry out the
non-perturbative numerical computation of the renormalization constants of $T_{\mu\nu}$~\cite{DGPinprep}.

Although the strategy is general,
for definiteness we consider the Wilson formulation of quarks on the lattice with and without the 
Sheikholeslami-Wohlert improving term~\cite{Sheikholeslami:1985ij}. In the presence of the latter, we also
carry out the analysis for the on-shell O($a$)-improvement of the energy-momentum tensor field.
The implementation of the renormalization and of the improvement program turn out to be 
greatly simplified because the thermal theory enjoys de-facto automatic O($a$)-improvement at
finite temperature.

The validity of the proposal is scrutinized analytically to 1-loop order in lattice perturbation theory
with shifted and twisted (for quarks only) boundary conditions. The results obtained this way serve also
to give a 1-loop improved definition of the non-perturbative renormalization constants with the aim of
reducing discretization effects.
For completeness, we notice that over the last few years alternative methods, based on the Yang--Mills
gradient flow \cite{Luscher:2010iy},
have also been explored for renormalizing non-perturbatively the energy-momentum
tensor~\cite{Suzuki:2013gza,DelDebbio:2013zaa,Makino:2014taa,Capponi:2015ahp,Taniguchi:2016ofw,Harlander:2018zpi}.  

The paper is organized as follows. In section~\ref{sec:ContinuumTheory} we derive the basic identities
which are later enforced on the lattice to fix the renormalization constants of the energy-momentum tensor,
and we construct the renormalization-group-invariant (RGI) definitions of its gluonic and fermionic components.
In the following section we show how these relations define non-perturbatively the energy-momentum tensor
on the lattice, while in section \ref{sec:Oaimp} we discuss the O($a$) improvement.
In the following section we investigate
the renormalization conditions in perturbation theory, and we compute the renormalization constants and the
improvement coefficients of the gluonic and fermionic parts of the energy-momentum tensor to 1-loop order.
Section~\ref{sec:concl} contains our conclusions. Notations, conventions, and technical details
are reported in several appendices.

\section{The energy-momentum tensor in the continuum\label{sec:ContinuumTheory}}
In this section we consider QCD in the continuum, for definitions and conventions
see appendices~\ref{App:conv} and \ref{App:cont}. We are interested in correlation functions of the
energy-momentum tensor $T_{\mu\nu}$ with gauge-invariant operators inserted at a physical distance. 
As reviewed in \cite{Caracciolo:1991cp,Giusti:2015daa}, it is appropriate in those cases to consider the
symmetric gauge-invariant definition of the energy-momentum tensor given by 
\begin{equation}
  \label{eq:Tmunu}
  T_{\mu\nu} = T^{G}_{\mu\nu} + T^{F}_{\mu\nu}\; ,
\end{equation}
where the first term is the gluonic component\footnote{Repeated indices are summed over unless explicitly specified.}
\begin{equation}
  \label{eq:TG}\displaystyle
  T^{G}_{\mu\nu} = {1 \over g_0^2}\left\{F^a_{\mu\alpha} F^a_{\nu\alpha}  
  - {1\over 4} \delta_{\mu\nu} F^a_{\alpha\beta} F^a_{\alpha\beta} \right\} \; , 
\end{equation}
while the second is the fermionic one given by
\be\label{eq:TF}
T^{F}_{\mu\nu} =   
{1 \over 4}\left\{\psibar\dirac\mu \overleftrightarrow{D}_\nu \psi +
  \psibar \dirac\nu \overleftrightarrow{D}_\mu\psi \right\}
 -\, {1 \over 4} \delta_{\mu\nu} \psibar \left\{{1 \over 2}\dirac\alpha \overleftrightarrow{D}_\alpha +M_0\right\}\psi \; ,
\ee
where $\overleftrightarrow{D}_\mu$ is defined in Eq.~(\ref{eq:Dbf}).

\subsection{Thermal QCD in the presence of shift and imaginary chemical potential}
For our strategy it is instrumental to consider the thermal theory with a non-zero imaginary chemical potential
formulated in a moving reference frame. Its grand canonical partition function reads
\be\label{eq:Z}
{\cal Z}(L_0,\vec\xi,\mu_{\cal I}) = \Tr\{e^{-L_0(\widehat H-i\vec\xi\cdot\widehat {\vec P} -i \mu_{\cal I} \widehat{N})}\}\; ,  
\ee
where the trace is over all the states of the Hilbert space, $L_0$ is the finite length of the
temporal direction, and $\widehat{N}$ is the quark number operator (three times the baryon number).
The Hamiltonian $\widehat H$,  the total momentum operator $\vec {\widehat P}$, and the imaginary
chemical potential $\mu_{\cal I}$ are expressed in a reference frame where the system is moving at a
velocity $\vec v$ which, by analytic continuation, we fix to the imaginary value $\vec v=i \vec\xi$.

In the Euclidean path-integral formalism the partition function (\ref{eq:Z}) is given by
\be
{\cal Z}(L_0,\vec\xi,\mu_{\cal I}) = Z(L_0,\bsxi,\theta_0) \Big|_{\theta_0 = -L_0\mu_{\cal I}}\; ,
\ee
where $Z(L_0,\bsxi,\theta_0)$ is the ordinary QCD path integral as defined in
Eq.~(\ref{eq:Zcont}) with fields which, in the time direction, satisfy periodic boundary
conditions up to a shift $L_0 \vec\xi$~\cite{Giusti:2010bb,Giusti:2011kt,Giusti:2012yj} and
a twist of the
fermion fields~\cite{Hasenfratz:1983ba}. Specifically, the gauge field $A_\mu$ obeys 
\begin{equation}
	\label{eq:Abcs}
	A_\mu(x_0+L_0,\bsx)= A_\mu(x_0,\bsx - L_0\bsxi)\; ,
\end{equation}
while the fermion fields, on top of the usual minus sign, pick up also a non-trivial twist phase
at the boundaries so that 
\ba
  \psi(x_0+L_0,\bsx) & = & -e^{i \theta_0}\, \psi(x_0,\bsx - L_0\bsxi)\; ,\nonumber\\[0.25cm]  
  \psibar(x_0+L_0,\bsx) & = & -e^{-i \theta_0}\, \psibar(x_0,\bsx - L_0\bsxi)\; .\label{eq:psibcs}
\ea
The free-energy density is given by
\begin{equation}
	\label{eq:free-energy}
	f(L_0,\bsxi,\theta_0) = -{1\over L_0 V} \ln Z(L_0,\bsxi,\theta_0)\; ,
\end{equation}
where $V=L_1 L_2 L_3$ is the spatial volume of the box. In the thermodynamic limit, which is
always assumed in this section, the invariance of the dynamics under the SO($4$) group
implies~\cite{Giusti:2012yj}
\begin{equation}\label{eq:bella1}
f(L_0,\bsxi,\theta_0) = f(L_0 \sqrt{1+\vec \xi^2},\boldsymbol{0},\theta_0)\; , 
\end{equation}
where the parameter $\theta_0$ remains the same on the two sides of the equation since the conserved
quark number charge is a relativistic invariant. The effect of the shift is therefore  
to lower the physical temperature and the chemical potential of the system
of the same factor, i.e. from $T=L^{-1}_0$ to $T=(L_0 \sqrt{1+\vec \xi^2})^{-1} $ and from $\mu_{\cal I}$  to
$\mu_{\cal I}/\sqrt{1+\vec \xi^2}$ respectively, with respect to the system with
the same values of $L_0$ and $\mu_{\cal I}$ but no shift. The entropy density reads
\be\displaystyle
{s \over T^3} = - L_0^4\, (1+\vec\xi^2)^2
\left\{ {(1+\vec\xi^2) \over \xi_k} \langle  T_{0k} \rangle_{\vec\xi,\theta_0} +
i \mu_{\cal I}\, \langle V_0 \rangle_{\vec\xi,\theta_0}  \right\}\; ,
\ee
where $V_0 = \bar\psi \gamma_0 \psi$. This  
formula generalizes the one in Ref.~\cite{Giusti:2012yj} to the
case of a non-zero chemical potential. The expectation values of the space-time components of the
energy-momentum
tensor dictate the dependence on $\vec \xi$ of the free-energy density
to be~\cite{Giusti:2010bb,Giusti:2011kt,Giusti:2012yj} 
\be
\langle T_{0k} \rangle_{\vec\xi,\theta_0} =  \displaystyle -
{\partial \over \partial \xi_k} f(L_0,\bsxi,\theta_0) \;,\label{eq:dxi1}\\[0.25cm]
\ee
while its $\theta_0$-dependence is determined by the average value of the temporal component of the
vector current so to satisfy
\ba
\langle V_0 \rangle_{\vec\xi,\theta_0} & = &  -i L_0 {\partial \over \partial \theta_0} f(L_0,\bsxi,\theta_0)\; .
\label{eq:dthet1}
\ea
As a result, the significant dependence of $\langle T_{0k} \rangle_{\vec\xi,\theta_0}$ on $\theta_0$
can be written as 
\be\displaystyle
\langle T_{0k} \rangle_{\vec\xi,\theta^A_0} - \langle T_{0k} \rangle_{\vec\xi,\theta^B_0} =
{i \over L_0} \int_{\theta^A_0}^{\theta^B_0} d \theta_0\, {\partial \over \partial \xi_k}
\langle V_0 \rangle_{\vec\xi,\theta_0}\; ,
\label{eq:bellissima}
\ee
a relation which turns out to be useful on the lattice.

The expectation values of the
traceless diagonal components of the energy-momentum tensor are related to those of
the space-time components via the Lorentz transformation discussed
in sections 2 and 4 of Ref.~\cite{Giusti:2012yj}, e.g. (no summation over repeated indices)
\ba
\langle T_{0k} \rangle_{\vec\xi,\theta_0} & = & {\xi_k \over 1-\xi_k^2} 
\left\{\langle T_{00} \rangle_{\vec\xi,\theta_0}  
- \langle T_{kk} \rangle_{\vec \xi,\theta_0}\,\right\}\; ,\label{eq:WIodd}\\[0.25cm]
\langle T_{0k} \rangle_{\vec\xi,\theta_0} & = & \xi_k 
\left\{\langle T_{00} \rangle_{\vec\xi,\theta_0}  
- \langle T_{jj} \rangle_{\vec \xi,\theta_0}\,\right\}\qquad (j\neq k, \xi_j=0 )\; .
\label{eq:WIodd2}
\ea
Such relations can be imposed on the lattice to fix their relative normalization.
Finally for the trace it
holds~\cite{Giusti:2015daa}
\be\label{eq:sing}
{\partial \over \partial \xi_k} \langle T_{\mu\mu} \rangle_{\vec\xi,\theta_0} =
{1 \over (1+\vec\xi^2)^2} {\partial \over \partial \xi_k}
\left[{(1+\vec\xi^2)^3 \over \xi_k} \langle T_{0k} \rangle_{\vec\xi,\theta_0} \right]\; . 
\ee
By combining Eq.~(\ref{eq:bellissima}) with Eqs.~(\ref{eq:WIodd})--(\ref{eq:sing}),
the $\theta_0$-dependence of the expectation values of the diagonal components of
$T_{\mu\nu}$ can be related to the one of the vector current as well.

\subsection{Finiteness of $T_{\mu\nu}$ and trace anomaly\label{sec:renoTc}}
In dimensional regularization the energy-momentum tensor (\ref{eq:Tmunu}) is decomposed as
\be
T_{\mu\nu} = \tau_{\mu\nu} + {1 \over 4} \delta_{\mu\nu} \bar \tau \; , 
\ee
with
\be
\tau_{\mu\nu} = \tau^{G}_{\mu\nu} + \tau^{F}_{\mu\nu}\; ,  
\ee
where
\bea
\tau^{G}_{\mu\nu} & = & {1 \over g_0^2}
\left\{F^a_{\mu\alpha} F^a_{\nu\alpha} - 
{1 \over D}\delta_{\mu\nu} F^a_{\alpha\beta} F^a_{\alpha\beta} \right\}\; ,\\[0.25cm]
\tau^{F}_{\mu\nu} & = &  {1 \over 4}\left\{\psibar\dirac\mu \overleftrightarrow{D}_\nu \psi +
\psibar \dirac\nu \overleftrightarrow{D}_\mu\psi \right\}
 -\, {1 \over D} \delta_{\mu\nu} \psibar \left\{\frac{1}{2}\dirac\alpha \overleftrightarrow{D}_\alpha\right\}\psi\; . 
 \eea
The singlet operator is
\be\label{eq:stark2}
\bar\tau = \bar\tau^{G} + \bar\tau^{F}\; ,  
\ee
where 
\be
\bar\tau^{G} =  {2 \epsilon \over D g_0^2} F^a_{\alpha\beta} F^a_{\alpha\beta}\; ,  
\ee
and up to terms that vanish by the equation of motion of
the fermion fields
\be
\bar\tau^{F} = - {4 \over D}\, \psibar M_0 \psi\;,   
\ee
with $D=4-2\epsilon$.

\subsubsection{The non-singlet\label{sec:non-sing}}
The dimension-four gauge-invariant fields $\tau^{G}_{\mu\nu}$ and $\tau^{F}_{\mu\nu}$ are parity and
charge conjugation invariant, and transform
as a two-index traceless symmetric irreducible representation of the SO($D$) group. No
other gauge-invariant field with the same quantum numbers and dimension $\leq 4$ can be constructed.
Since the derivative of the free-energy density with respect to the shift is finite once the
bare parameters of the theory have been renormalized~\cite{Giusti:2010bb,Giusti:2011kt,Giusti:2012yj},
i.e. Eq.~(\ref{eq:dxi1}), we may choose the renormalization pattern to be 
\be
\left(\begin{array}{c}
\tau^{G}_{\mu\nu}\\[0.25cm]
\tau^{F}_{\mu\nu}
\end{array}\right) = {\cal Z}_\tau
\left(\begin{array}{c}
\tau^{G, R}_{\mu\nu}\\[0.25cm]
\tau^{F, R}_{\mu\nu}
\end{array}\right)
\; , \qquad
{\cal Z}_\tau =
\left(\begin{array}{cc}
1+ z_{g} & - z_{f}\\[0.25cm]
- z_{g} & 1+ z_{f}  
\end{array}\right)\; , 
\ee
where the superscript $R$ labels renormalized quantities in the desired scheme.
The anomalous-dimension matrix takes the form  
\be
\gamma_\tau(g) = -
{1\over {\cal Z}_\tau}\mu {d\over d \mu} {\cal Z}_\tau
=
\left(\begin{array}{cc}
a & c\\[0.25cm]
- a & - c  
\end{array}\right)\; = g^2 \sum_{k=0}^{\infty} (\gamma_\tau)_{k}\, g^{2k}\; ,
\ee
where at leading order
in $g^2$ the coefficient matrix is 
\be
(\gamma_\tau)_0 = \left(\begin{array}{cc}
a_0 & c_0\\[0.25cm]
- a_0 & - c_0  
\end{array}\right)
=R_0^{-1}
\left(\begin{array}{cc}
\gamma_0 & 0\\[0.25cm]
   0     & 0
\end{array}\right)
R_0\; , \qquad
R_0 =
\left(\begin{array}{cc}
\displaystyle {2\, a_0\over \gamma_0} & \displaystyle {2\, c_0 \over \gamma_0}\\[0.25cm] 
1 & 1
\end{array}\right)\; ,
\ee
with~\cite{Bardeen:1978yd,Capitani:1994qn}
\be
a_0=- {1 \over (4\pi)^2} {4 N_f \over 3}\; , \quad
c_0={1 \over (4 \pi)^2} {8 \over 3} {N_c^2-1 \over N_c}\; , \quad
\gamma_0 = (a_0 - c_0)\; . 
\ee
It is then straightforward to verify that
\be
\gamma_\tau = R^{-1}_0
\left(\begin{array}{cc}
\gamma & \alpha\\[0.25cm]
0 & 0 
\end{array}\right)
R_0\; , 
\ee
where 
\be
\alpha = 2\, {c a_0 - a c_0 \over \gamma_0 } = g^4 \sum_{k=0}^{\infty} \alpha_{k+1}\, g^{2k}\; , \qquad
\gamma = a-c = g^2 \sum_{k=0}^{\infty} \gamma_{k}\, g^{2k}\; .
\ee
By following Refs.~\cite{Capitani:1998mq,Giusti:2004an,Giusti:2002lgml}, we can define the
RGI gluonic and fermionic components of the energy-momentum tensor as
\be
\left(\begin{array}{c}
\tau^{G,{\rm RGI}}_{\mu\nu}\\[0.25cm]
\tau^{F,{\rm RGI}}_{\mu\nu}
\end{array}\right) =
\Theta_\tau(g)
\left(\begin{array}{c}
\tau^{G, R}_{\mu\nu}\\[0.25cm]
\tau^{F, R}_{\mu\nu}
\end{array}\right)\; ,
\ee
where $\Theta_\tau(g)$ satisfies the differential equation 
\be
\beta(g) {\partial\over\partial g} \Theta_\tau(g) + \Theta_\tau(g) \gamma_\tau(g) =  0\; 
\ee
with the initial condition\footnote{For $\gamma_0/b_0>-2$, e.g. $N_c=3$ and $N_f\leq 5$, the subtraction proportional
to $\alpha_1$ in Eq.~(\ref{eq:RGIsinglim}) is harmless. If $\gamma_0/b_0 \leq -4$, e.g. $N_c=3$ and $N_f> 9$,
further subtractions are required.}
\be\label{eq:RGIsinglim}
\lim_{g\rightarrow 0}\Bigg\{\Theta_\tau(g) (2 b_0 g^2)^{-{(\gamma_\tau)_{0}\over 2 b_0}}
- {\alpha_1 g^2\over 2 (2 b_0+\gamma_0)} (2 b_0 g^2)^{{\gamma_{0}\over 2 b_0}}
\left(\begin{array}{cc}
 1 & 1 \\[0.25cm]
-1 &-1 
\end{array}\right)
\Bigg\} =  1\!\!1\; , 
\ee
and with $b_0$ being the first coefficient of the $\beta$-function defined in
Eq.~(\ref{eq:bo1}).
The solution has the same structure as ${\cal Z}_\tau$, and it can be written as
\be\label{eq:thetatau}
\Theta_\tau(g) = R^{-1}_0
\left(\begin{array}{cc}
\theta_1 & \theta_2 \\[0.25cm]
0 & 1 
\end{array}\right)
R_0\; ,
\ee
where
\ba
\theta_1(g) & = & (2 b_0 g^2)^{{\gamma_{0}\over 2 b_0}} \exp\Big\{- \int_0^{g}
\left[{\gamma(\bar g)\over \beta(\bar g)} + {\gamma_0 \over b_0 \bar g}\right]d \bar g \Big\}\; ,\\[0.25cm]
\theta_2(g) & = & {\alpha_1 g^2\over 2 b_0+\gamma_0} (2 b_0 g^2)^{{\gamma_{0}\over 2 b_0}}
-\int_0^{g} \left[ {\alpha(\bar g)\over \beta(\bar g)} \theta_1(\bar g) 
+ {\alpha_1 \bar g \over b_0} (2 b_0 \bar g^2)^{{\gamma_{0}\over 2 b_0}}
\right] d \bar g\; , \label{eq:theta2}
\ea
and the $\beta$-function $\beta(g)$ is defined in Eq.~(\ref{eq:betagsol}). 
From these equations it follows that
\ba
\tau_{\mu\nu} & =  & \tau^{G,{\rm RGI}}_{\mu\nu} + \tau^{F,{\rm RGI}}_{\mu\nu}\;,\\
t_{\mu\nu} & = & \tau_{\mu\nu}^{G,{\rm RGI}}-\tau_{\mu\nu}^{F,{\rm RGI}}
\ea
are two independent RGI fields. Once defined, they allow for an unambiguous scale- and scheme-independent
separation of the energy-momentum tensor in two parts, and therefore of its form factors among
physical states, which tend to the free gluonic and fermionic contributions when
$g\rightarrow 0$. At finite temperature,
for instance, this allows for an unambiguous split of the entropy density in two
parts which tend, in the infinite temperature limit, to the Stefan-Boltzmann values
for gluons and fermions respectively.

\subsubsection{The singlet}
The gluonic and fermionic components of $\bar\tau$ are singlets under SO($D$), and
therefore they mix with the identity. A natural prescription for subtracting
this contribution leads to the renormalization pattern
\be\label{eq:renops}
\left(\begin{array}{c}\displaystyle
{1 \over g_0^2} F^a_{\alpha\beta} F^a_{\alpha\beta} -
\Big\langle {1 \over g_0^2} F^a_{\alpha\beta} F^a_{\alpha\beta} \Big\rangle_0\\[0.5cm]
\!\!\!\!\!\!\!\!\!\! \psibar M_0 \psi - \Big\langle \psibar M_0 \psi \Big\rangle_0
\end{array}\right) =
{\cal Z}_{\bar\tau}
\left(\begin{array}{c}\displaystyle
\left\{{1 \over g_0^2} F^a_{\alpha\beta} F^a_{\alpha\beta}\right\}^{R}\\[0.5cm]
\left\{ \psibar M_0 \psi \right\}^{R}
\end{array}\right)\; , 
\ee
where $\langle \dots \rangle_0$ indicates the expectation value 
for $L_0\rightarrow \infty$ (zero temperature), and 
\be
{\cal Z}_{\bar\tau} =
\left(\begin{array}{cc}
1+ \bar z_{g} & \bar z_{f}\\[0.25cm]
 0  & 1   
\end{array}\right)\; .
\ee
Also for the singlet, we can define the RGI gluonic and fermionic operators as
\be\label{eq:RGIsing0}
\left(\begin{array}{c}\displaystyle
\left\{{1 \over g_0^2} F^a_{\alpha\beta} F^a_{\alpha\beta}\right\}^{\rm RGI}   \\[0.5cm]
\left\{ \psibar M_0 \psi \right\}^{\rm RGI}
\end{array}\right) =
\Theta_{\bar\tau}(g)
\left(\begin{array}{c}\displaystyle
\left\{{1 \over g_0^2} F^a_{\alpha\beta} F^a_{\alpha\beta}\right\}^{R} \\[0.5cm]
\left\{ \psibar M_0 \psi \right\}^{R}
\end{array}\right)\; ,
\ee
where the $2\times 2$ matrix $\Theta_{\bar\tau}(g)$ satisfies the differential
equation 
\be\label{eq:RGIsing}
\beta(g) {\partial \over \partial g} \Theta_{\bar\tau}(g) + \Theta_{\bar\tau}(g) \gamma_{\bar\tau}(g) = 0\;  
\ee
with the initial condition
\be\label{eq:RGIsinginit}
\lim_{g\rightarrow 0}\Big\{ \Theta_{\bar\tau}(g) (2 b_0 g^2)^{{-(\gamma_{\bar\tau})_0 \over 2 b_0}} \Big\} = 1\!\!1\; .
\ee
The anomalous-dimension matrix is given by
\be
\gamma_{\bar\tau}(g) = - {1\over {\cal Z}_{\bar\tau}}\mu {d\over d \mu} {\cal Z}_{\bar\tau} =  
\left(\begin{array}{cc}\displaystyle
\bar\gamma & \bar\alpha \\[0.25cm]
0 & 0  
\end{array}\right) = g^2 \sum_{k=0}^{\infty} (\gamma_{\bar\tau})_k \, g^{2k}\; ,
\ee
where at leading order in $g^2$, see below, the coefficient matrix is
\be
(\gamma_{\bar\tau})_0 =
\left(\begin{array}{cc}\displaystyle
2 b_0 & -4 d_0 \\[0.25cm]
0 & 0  
\end{array}\right)\; . 
\ee
The solution of Eqs.~(\ref{eq:RGIsing}) and (\ref{eq:RGIsinginit}), analogous to the one in
Eqs.~(\ref{eq:thetatau})--(\ref{eq:theta2}), reads
\be\label{eq:thetataubar}
\Theta_{\bar \tau}(g) =
\left(\begin{array}{cc}
\bar\theta_1 & \bar\theta_2 \\[0.25cm]
0 & 1 
\end{array}\right)\; ,
\ee
where
\ba
\bar\theta_1(g) & = & (2 b_0 g^2) \exp\Big\{- \int_0^{g}
\left[{\bar\gamma(\bar g)\over \beta(\bar g)} + {2 \over \bar g}\right]d \bar g \Big\}\; ,\\[0.25cm]
\bar\theta_2(g) & = & {2 d_0 \over b_0} 
-\int_0^{g} {\bar\alpha(\bar g)\over \beta(\bar g)} \bar \theta_1(\bar g) d \bar g\; .
\label{eq:theta_sing} 
\ea
In dimensional regularization, by deriving Eq.~(\ref{eq:dxi1}) with respect to the renormalized
coupling at fixed renormalized quark mass $M$, see 
Eqs.~(\ref{eq:free-energy}), (\ref{eq:renDRg}) and (\ref{eq:renDRm}), we obtain
\be\label{eq:stark}
2 g {\partial\over\partial g} \big\langle T_{0k} \big\rangle_{\vec\xi,\theta_0} = - {\epsilon g\over \beta(\epsilon,g)}
{\partial \over \partial \xi_k} \Big\langle {1 \over g_0^2} F^a_{\alpha\beta} F^a_{\alpha\beta}\Big\rangle_{\vec\xi,\theta_0}
+ {2 g \gamma_m\over \beta(\epsilon,g)} {\partial \over \partial \xi_k}
\big\langle \psibar M_0 \psi \big\rangle_{\vec\xi,\theta_0}\; ,
\ee
where $\gamma_m$ is the anomalous dimension of the quark mass defined in Eq.~(\ref{eq:gam_m}). Since
the l.h.s. of (\ref{eq:stark}) is finite, the r.h.s. defines the renormalized
gluonic operator in Eq.~(\ref{eq:renops}) with the choice 
\be
\bar z_{g} = - {\beta(g) \over\epsilon g}\; ,\qquad
\bar z_{f} = {2 \gamma_m \over \epsilon}\; . 
\ee
Analogously by deriving with respect to the renormalized quark masses,
it follows that we can take $({\cal Z}_{\bar \tau})_{22}=1$. These values of the
renormalization constants correspond exactly to the MS prescription,
and therefore the superscript $R$ in Eq.~(\ref{eq:renops}) labels
renormalized fields defined in the MS scheme in this case. The anomalous dimension matrix reads
\be
\gamma_{\bar\tau}(g) =  -\beta(\epsilon,g) {1\over {\cal Z}_{\bar\tau}}
{\partial\over \partial g} {\cal Z}_{\bar\tau}  =
g {\partial\over \partial g} 
\left(\begin{array}{cc}\displaystyle
-{\beta(g)\over g} & 2 \gamma_m \\[0.25cm]
0 & 0  
\end{array}\right)\; .
\ee
By inserting these findings in Eq.~(\ref{eq:stark2}), and taking the limit
$\epsilon\rightarrow 0$, one obtains the
result~\cite{Adler:1976zt,Collins:1976yq,Tarrach:1981bi}
\be\label{eq:Tran2}
\bar \tau - \langle \bar \tau \rangle_0 = 
- {\beta(g) \over 2 g} \left\{{1 \over g_0^2} F^a_{\alpha\beta} F^a_{\alpha\beta}\right\}^{R}
- (1-\gamma_m) \left\{ \psibar M_0 \psi \right\}^{R}\; ,
\ee
which is valid to all orders in perturbation theory. 
By noticing that the solution of Eqs.~(\ref{eq:RGIsing}) and (\ref{eq:RGIsinginit}) is
\be
\Theta_{\bar\tau}(g) =\left(\begin{array}{cc}\displaystyle
- {2 \beta(g) \over g} & \displaystyle \left[{2 d_0\over b_0} + 4 \gamma_m\right] \\[0.5cm]
0 & 1  
\end{array}\right)\; ,
\ee
we can finally rewrite Eq.~(\ref{eq:Tran2}) for
the trace anomaly in terms of RGI operators as 
\be
\bar\tau - \langle \bar\tau \rangle_0 = {1\over 4} \left\{{1 \over g_0^2} F^a_{\alpha\beta} F^a_{\alpha\beta}\right\}^{\rm RGI}
- \big(1 + {d_0\over 2 b_0}\big) \left\{ \psibar M_0 \psi \right\}^{\rm RGI}\; .               
\ee
Analogously to the non-singlet case, we can write
\ba
\bar\tau - \langle \bar\tau \rangle_0  = \bar\tau^{G,{\rm RGI}} + \bar\tau^{F,{\rm RGI}}\; ,
\ea
where
\ba
\bar\tau^{G,{\rm RGI}} & = & {1\over 4} \left\{{1 \over g_0^2} F^a_{\alpha\beta} F^a_{\alpha\beta}\right\}^{\rm RGI} - {d_0\over 2 b_0}
                           \left\{ \psibar M_0 \psi \right\}^{\rm RGI}\\[0.25cm]
\bar\tau^{F,{\rm RGI}} & = &  - \left\{ \psibar M_0 \psi \right\}^{\rm RGI}\; . 
\ea
Considerations analogous to those made for the non-singlet components of the energy-momentum
tensor apply also in this case. 

\subsection{Finite volume}
In a finite volume the theory needs to be further specified by the boundary conditions imposed
on the fields in the spatial directions. For the gauge field we supplement
Eq.~(\ref{eq:Abcs}) by standard periodic boundary conditions in the spatial directions, 
\begin{equation}
	\label{eq:Abcs_finiteV}
	A_\mu(x_0,\bsx+\hat{k} L_k)= A_\mu(x_0,\bsx)\; ,
\end{equation}
where $\hat{k}$ is the unit vector in the direction $k$. Note that this implies
$-L_k/2 \leq L_0 \xi_k < L_k/2$. For quark and anti-quark fields the boundary conditions
in Eqs.~(\ref{eq:psibcs}) are supplemented by periodic boundary conditions in the spatial
directions up to a non-trivial twist phase $\theta_k$~\cite{Luscher:1996sc}
\ba
\psi(x_0,\bsx+\hat{k} L_k) & = & e^{i \theta_k}\, \, \psi(x_0,\bsx)\; ,\nonumber\\[0.25cm]
\psibar(x_0,\bsx+\hat{k} L_k) & = &  e^{-i\theta_k}\, \psibar(x_0,\bsx)\; .\label{eq:psibcs_finiteV}
\ea
The finite length of the spatial directions and the angles $\theta_k$ softly break the SO($4$) group,
and $\theta_k$ breaks charge conjugation as well. As a consequence
Eqs.~(\ref{eq:WIodd})--(\ref{eq:sing}) are modified in a finite volume.

For instance, Eq.~(\ref{eq:WIodd}) needs to be replaced by Eq.~(4.25) of
Ref.~\cite{Giusti:2012yj}. In the latter there is an extra term proportional to the expectation value of
$T_{0k}$ computed in the reference frame where the velocity in direction $k$ is null while the other
velocity components are unchanged. For this extra term to be null, the space-time geometry
must satisfy the condition~\cite{Giusti:2012yj}
\be
\label{eq:PeriodicCondition}
{L_k \xi_k \over L_0 (1 +\xi_k^2)} =q\in\mathbb{Z}\; ,
\ee
together with
\be
b_0 =  \vec\xi\cdot\vec b - \xi_k b_k  \qquad \mbox{or}\qquad b_0 =  \vec\xi\cdot\vec b + {b_k \over \xi_k}\; , 
\ee
where $b_\mu = \theta_\mu/L_\mu$. Similarly Eq.~(\ref{eq:WIodd2}) picks up two extra terms: one proportional to
the expectation value of $T_{0k}$ as before, and a second one proportional to the expectation value of $(T_{kk} - T_{jj})$.
For both these terms to be null in the reference frame where the velocity in direction $k$ is null while
the other components are unchanged, the condition (\ref{eq:PeriodicCondition}) and
$L_k/\sqrt{1+\xi_k^2}=L_j$ have to be met together with
\be\label{eq:allb}
\left\{\begin{array}{rcl}
b_0 & = & \vec\xi\cdot\vec b - \xi_k b_k \\[0.25cm]
b_k & = & {L_j \over L_k} b_j  
\end{array}\right. \qquad \mbox{or} \qquad
\left\{\begin{array}{rcl}
b_0 & = &  \vec\xi\cdot\vec b + {b_k \over \xi_k}\\[0.25cm]
b_j & = & 0  
\end{array}\right.\; .
\ee
In the simplified case in which the only non-null component of the shift is $\xi_k$, 
the conditions in Eqs.~(\ref{eq:PeriodicCondition})--(\ref{eq:allb}) can be summarized as
\be\label{eq:allb1}
{L_k \xi_k \over L_0 (1 +\xi_k^2)} =q \; ,\qquad 
\left\{\begin{array}{rcl}
\theta_0 & = & 0 \\[0.25cm]
\theta_k & = & \theta_j  
\end{array}\right. \qquad \mbox{or} \qquad
\left\{\begin{array}{rcl}
\theta_k & = &  q \theta_0 \\[0.25cm]
\theta_j & = & 0  
\end{array}\right.\; .
\ee
It must be said, however, that in the presence of a mass gap the
extra terms acquired by Eqs.~(\ref{eq:WIodd})--(\ref{eq:sing}) vanish exponentially with the
length of the spatial directions, and therefore they are expected to be negligible in large enough volumes.

\section{The energy-momentum tensor on the lattice\label{sec:LatQCD}}
A non-perturbative definition of the theory presented in Sect.~\ref{sec:ContinuumTheory} is achieved
by introducing a four-dimensional Euclidean hypercubic lattice of spacing $a$ which acts as an
ultraviolet regulator. The gauge potential $A_\mu$ is replaced as usual by the
$\SUn$-valued gauge fields $U_\mu$ residing on the links of the lattice, while the quark and anti-quark fields
$\psi,\psibar$ are defined on the sites\footnote{We use the same notation for lattice and continuum quantities,
since any ambiguity is resolved from the context. As usual, the continuum limit value of a renormalized
lattice quantity, identified with the superscript $R$, is the one to be identified with its
continuum counterpart.}. Although the non-perturbative renormalization strategy presented in this paper is general,
for definiteness we consider the Wilson formulation for gluons and quarks, see
appendix~\ref{App:lat} for the details.

A discretization of the energy-momentum tensor is obtained by replacing the fields and the derivatives appearing in the
continuum expressions (\ref{eq:Tmunu})--(\ref{eq:TF}) with their lattice counterparts. In particular we
define~\cite{Caracciolo:1989pt,Caracciolo:1989bu,Caracciolo:1991cp,Caracciolo:1991vc} 
\begin{equation}
  \label{eq:BareEMT}
  {T}_{\mu\nu} = {T}^G_{\mu\nu} + {T}^F_{\mu\nu}\; , 
\end{equation}
with the gluonic component given by
\begin{equation}
  \label{eq:BareTG}\displaystyle
  T^G_{\mu\nu} = {1 \over g_0^2}\left\{F^a_{\mu\alpha} F^a_{\nu\alpha}  
  - {1\over 4} \delta_{\mu\nu} F^a_{\alpha\beta} F^a_{\alpha\beta} \right\} \; , 
\end{equation}
where
\be\label{eq:FmunuL}
{F}_{\mu\nu}^a=2\,\Tr\{\widehat{F}_{\mu\nu} T^a\}\; ,\qquad  {F}_{\mu\nu} = {F}_{\mu\nu}^a T^a\; , 
\ee
and with $\widehat{F}_{\mu\nu}$ being the clover discretization of the field
strength tensor in Eq.~(\ref{eq:CloverFmunu}) which, at variance with
${F}_{\mu\nu}$, is not traceless.
For the fermionic part we take
\ba
  \label{eq:BareTF}
{T}^F_{\mu\nu} & = &  {1 \over 8}\bigg\{
\psibar \dirac\mu\big[\nabdbarstar\nu+\nabdbar\nu\big]\psi +
\psibar \dirac\nu\big[\nabdbarstar\mu+\nabdbar\mu\big]\psi \bigg\}\nonumber\\[0.25cm]
&  & -\, {1 \over 4} \delta_{\mu\nu} \psibar
\bigg\{{1 \over 4}\dirac\alpha\Big(\nabdbarstar\alpha+\nabdbar\alpha\Big)+M_0\bigg\}\psi\; ,
\ea
where $\nabdbar\mu$ and $\nabdbarstar\mu$ are defined in Eq.~(\ref{eq:DbfL}). 
In the naive continuum limit $a\rightarrow 0$, these expressions tend to the continuum ones in
Eqs.~(\ref{eq:TG}) and (\ref{eq:TF}). As 
any other discretization of ${T}_{\mu\nu}$, however, they need to be properly renormalized to guarantee that
their correlation functions satisfy the continuum Ward identities, up to discretization errors which
vanish in the continuum limit.

The target energy-momentum tensor in the continuum is a gauge-invariant operator of dimension 4,
which is a combination of a traceless two-index symmetric and a singlet 
irreducible representations of SO($4$) even under parity and charge conjugation. Since on the lattice
the SO($4$) symmetry reduces to the hypercubic group SW$_4$, the traceless
two-index symmetric representation splits into a sextet (non-diagonal components) and a triplet
(diagonal traceless components). At finite lattice spacing, the energy-momentum
tensor is thus a combination of gauge-invariant operators of dimension $d\leq 4$ 
which, under the hypercubic group, transform as one of those two representations and the
singlet~\cite{Caracciolo:1989pt}.
In QCD there are seven such operators plus the identity. We can take as a basis
(no summation over the repeated indices $\mu$ and $\nu$ in
Eqs.~(\ref{eq:TG3}) and (\ref{eq:TF3}))
\ba
{T}_{\mu\nu}^{G,\{6\}} & = & \big(1-\delta_{\mu\nu})
  {1\over g_0^2}\Big\{{F}^a_{\mu\alpha} {F}^a_{\nu\alpha} \Big\}\; ,\label{eq:TG6}\\[0.25cm]
  {T}_{\mu\nu}^{G,\{3\}} & = & {1\over g_0^2}\Big\{{F}^a_{\mu\alpha} {F}^a_{\mu\alpha} - {F}^a_{\nu\alpha} {F}^a_{\nu\alpha}\Big\}\; ,
  \label{eq:TG3}\\[0.25cm]
  {T}_{\mu\nu}^{G,\{1\}} & = &   \delta_{\mu\nu} {1\over 4 g_0^2}  F^a_{\alpha\beta} F^a_{\alpha\beta}\; ,
\ea
for the purely gluonic fields and
\ba
{T}_{\mu\nu}^{F,\{6\}} & = &  (1-\delta_{\mu\nu})
  {1 \over 8}\left\{\psibar\dirac\mu\big[\nabdbarstar\nu+\nabdbar\nu\big]\psi +
  \psibar\dirac\nu\big[\nabdbarstar\mu+\nabdbar\mu\big]\psi \right\}\; ,\label{eq:TF6}\\[0.125cm]
{T}_{\mu\nu}^{F,\{3\}} & = & {1 \over 4}\Big\{\psibar\dirac\mu\left[\nabdbarstar\mu+\nabdbar\mu\right]\psi
- \psibar\dirac\nu\left[\nabdbarstar\nu+\nabdbar\nu\right]\psi\Big\}\; , \label{eq:TF3}\\[0.125cm]
{T}_{\mu\nu}^{F,\{1a\}} & = &  \delta_{\mu\nu} {1 \over 16}\psibar \Big\{\dirac\alpha\Big(\nabdbarstar\alpha+\nabdbar\alpha\Big)\Big\}\psi
\; ,\qquad \\[0.125cm]
{T}_{\mu\nu}^{F,\{1b\}} & = & \delta_{\mu\nu} \psibar\psi\; ,\label{eq:TF1b}
\ea
for the fermionic ones\footnote{Notice that, at variance with the continuum, the operator ${T}_{\mu\nu}^{F,\{1b\}}$  is not a chiral
singlet, i.e. multiplied by the quark mass, since chiral symmetry is explicitly broken for Wilson fermions.}.
Since the hypercubic group is an exact symmetry of the lattice theory, a given field in
Eqs.~(\ref{eq:TG6})--(\ref{eq:TF1b}) can mix under renormalization only with those in the same
irreducible representation.

In this paper we focus on the definition of the sextet and triplet whose  
renormalization pattern is given by
\be
{T}_{\mu\nu}^{R,\{i\}} =  Z^{\{i\}}_{G}(g_0^2)\, {T}_{\mu\nu}^{G,\{i\}} + Z^{\{i\}}_{F}(g_0^2)\, {T}_{\mu\nu}^{F,\{i\}}\; ,
\quad i=6,3\; , \label{eq:SextetEMT}
\ee
where the renormalization constants $Z^{\{i\}}_{G}$ and $Z^{\{i\}}_{F}$ are finite and depend on $g_0^2$ only
since in the continuum the nonet component of the energy-momentum tensor has vanishing anomalous dimension.
In a regularization which breaks chiral symmetry explicitly, 
the renormalization pattern of the singlet component requires a rather different and more involved analysis which
fills a different publication by itself. Departure from scale invariance can, however, be extracted directly
from the Callan--Symanzik renormalization group equations or, for instance, from the r.h.s of Eq.~(\ref{eq:sing}).
Finally we note that, as anticipated in section \ref{sec:non-sing}, the determination of the
renormalized fields
\begin{equation}
    \label{eq:tmn}
    t^{R, \{i\}}_{\mu\nu}(\mu)=z_G^{\{i\}}(g_0^2,a\mu)\, T_{\mu\nu}^{G,\{i\}}
    -z_F^{\{i\}}(g_0^2,a\mu)\, T_{\mu\nu}^{F,\{i\}}\,,
    \quad
    i=6,3\,,
\end{equation}
and in particular of their RGI counterparts, gives access to the RGI gluonic and fermionic components
of the energy-momentum tensor. While the renormalization of $T^{\{i\}}_{\mu\nu}$ is fixed
by WIs, that of (\ref{eq:tmn}) can be obtained, for instance, by imposing suitable conditions on
its expectation values in the presence of shifted and twisted boundary conditions. A detail investigation
of this problem is in progress.

\subsection{Non-perturbative renormalization conditions\label{eq:secNP}}
The renormalization constants $Z^{\{i\}}_{G}$ and $Z^{\{i\}}_{F}$ can be determined
non-perturbatively by enforcing on the lattice the relations (\ref{eq:dxi1}) and (\ref{eq:WIodd})
up to discretization effects which vanish in the continuum limit. To this aim, shifted
boundary conditions for the links are 
\begin{equation}
  \label{eq:shift_lat}
  U_\mu(x_0+L_0,\bsx)= U_\mu(x_0,\bsx-L_0\bsxi)\; ,
  \qquad
  U_\mu(x_0,\bsx+\hat{k}L_k)= U_\mu(x_0,\bsx)\; ,
\end{equation}
while Eqs.~(\ref{eq:psibcs}) and (\ref{eq:psibcs_finiteV}) fix 
the boundary conditions for quark and anti-quark fields on the lattice too,
where the components of $\bsxi$ are discretized in integer units of $a/L_0$.
As very commonly adopted in lattice QCD, we opt for a mass-independent renormalization scheme.
The renormalized coupling $g_R$ and the renormalized quark mass $m_R$ for each
given flavour are thus related to the bare parameters as
\be
g_R^2 = Z_{g}(g^2_0,a\mu)\, g_0^2\, ,
\qquad m_R = Z_{m}(g^2_0,a\mu)\, m_{q}\, ,
\label{eq:gmR}
\ee
where $\mu$ is a renormalization scale, $m_q=m_0-m_{c}(g^2_0)$ is the subtracted mass, and 
$m_{c}(g^2_0)$ is the critical mass.

The two renormalization constants of the sextet are fixed by requiring that,
for two different set of values $\theta^A$ and $\theta^B$, where $\theta=(\theta_0,{\vec\theta})$, it holds
\begin{equation}
  \label{eq:LatticeWI1}
  \langle {T}^{R,\{6\}}_{0k} \rangle_{\bsxi,\theta} 
  = - \frac{\displaystyle \Delta {f}(L_0,\bsxi,\theta)}{\displaystyle \Delta \xi_k}\; ,
\end{equation}
where
\begin{equation}
  \label{eq:LatticeXiDerivative}
  \frac{\displaystyle \Delta {f}(L_0,\bsxi,\theta)}{\displaystyle \Delta \xi_k} = 
	{L_0\over 2a}\Big[{f}(L_0,\bsxi+\frac{a}{L_0} \hat{k},\theta)
	-{f}(L_0,\bsxi-\frac{a}{L_0} \hat{k},\theta)\Big]
\end{equation}
is a symmetric discrete approximation of the derivative of the free energy with respect to the $k$-th component
of the shift.
From a practical point of view, it is useful to combine Eq.~(\ref{eq:LatticeWI1}) for one set of values of the angles,
e.g. $\theta=\theta^A$, with the lattice realization of the identity (\ref{eq:bellissima})
\be\displaystyle
\langle {T}^{R,\{6\}}_{0k} \rangle_{\vec\xi,\theta^A} - \langle {T}^{R,\{6\}}_{0k} \rangle_{\vec\xi,\theta^B} =
{i \over L_0} \int_{\theta^A_0}^{\theta^B_0} d \theta_0\,
\frac{\displaystyle \Delta \langle V^R_0 \rangle_{\vec\xi,\theta}}{\displaystyle \Delta \xi_k}\; ,
\qquad \vec\theta^A = \vec\theta^B\; , 
\label{eq:bellissimaL}
\ee
since the r.h.s. can be computed very efficiently by Monte Carlo simulations. Moreover, if one takes
the temporal component of the flavour-singlet conserved lattice vector current
\begin{equation}
  \label{eq:VectorCurrent}
        V^c_\mu(x)= \frac{1}{2} \big\{\psibar(x+ a \hat{\mu}) U_\mu^\dag(x)(\gamma_\mu+1)\psi(x) +
                                          \psibar(x) U_\mu(x)(\gamma_\mu-1)\psi(x+ a \hat{\mu})\big\} \; ,
\end{equation}
then
\begin{equation}
V_\mu^R(x) = V^c_\mu(x)\; , 
\end{equation}
and the expectation value of the renormalized current on the r.h.s. of Eq.~(\ref{eq:bellissimaL}) corresponds to
the bare one\footnote{It is interesting to notice that the renormalization constant and the improvement coefficients
of the flavour-singlet lattice vector currents can be determined by comparing $\langle V^c_0 \rangle_{\vec\xi,\theta}$
with the analogous one for the local current and similarly for higher-point correlation functions.}.

Once the sextet renormalization constants have been fixed, those of the triplet can
be determined by enforcing on the lattice the analogous of Eqs~(\ref{eq:WIodd}) or (\ref{eq:WIodd2})
\bea
  \label{eq:LatticeWI3}
  \langle {T}^{R,\{6\}}_{0k} \rangle_{\bsxi,\theta}& = &{\xi_k\over 1-\xi_k^2}\,
  \langle {T}^{R,\{3\}}_{0k} \rangle_{\vec\xi,\theta}\\[0.125cm]
  \langle {T}^{R,\{6\}}_{0k} \rangle_{\vec\xi,\theta} & = &  \xi_k\, \langle {T}^{R,\{3\}}_{0j} \rangle_{\vec\xi,\theta}\; ,
  \qquad (j\neq k, \xi_j=0 )\; .
\label{eq:LatticeWI4}
\eea
for two different set of values $\theta=\theta^A,\theta^B$ of the twist angles. In a finite volume these
relations are satisfied only up to exponentially small finite-size effects, unless lattice sizes and twist
phases according to the
constraints given in Eqs.~(\ref{eq:PeriodicCondition})--(\ref{eq:allb}) are considered. The above renormalization conditions
are imposed at zero quark masses, i.e. all bare masses are set to the critical value $m_0=m_{cr}(g_0)$. In practice
this is possible thanks to the presence of a spectral gap in the lattice Dirac operator at finite temperature.

Finally, it is important to emphasize that the above renormalization
conditions are valid non-perturbatively, and are designed to be accessible to numerical Monte Carlo
computations. The Eq.~(\ref{eq:LatticeWI1}) can be studied numerically with a strategy analogous
to the one already successfully implemented for the Yang--Mills theory \cite{Giusti:2015daa}, while
Eqs.~(\ref{eq:bellissimaL}) and (\ref{eq:LatticeWI3}) require standard numerical techniques.
In this sense these conditions provide a practical strategy to define non-perturbatively
the energy-momentum tensor in QCD.

\section{O($a$)-improvement\label{sec:Oaimp}}
The Symanzik improvement programme has the purpose of accelerating the approach to
the continuum limit of field correlators. It is achieved by adding suitable counterterms to
the lattice action and to the fields multiplied by numerical coefficients which are properly
adjusted so to cancel discretization errors order by order in the lattice
spacing~\cite{Symanzik:1983dc,Symanzik:1983gh}. In the following 
we discuss how to implement this programme to O($a$)-improve on-shell matrix elements of the
sextet and triplet components of the energy-momentum tensor. For the clarity of the presentation,
we discuss separately the massless, mass-degenerate, and mass non-degenerate cases.

\subsection{Massless quarks}\label{Oa_massless}
The first step consists in identifying complete bases of dimension-5 gauge-invariant fields which, in the Symanzik
effective continuum theory, are parity and charge conjugation invariant and transform as sextets and triplets
under the SO($4$) hypercubic subgroup.
When all quarks are massless, i.e. $m_0 = m_{c}(g_0)$, there are no operators made of the
gauge field only. Complete bases of O($a$)-counterterms are built by 
projecting the fields
\begin{align}\label{eq:O1}
\psibar \sigma_{\mu\rho}\,F_{\nu\rho} \psi \, ,\qquad
\drv\rho\Big\{\psibar\sigma_{\mu\rho}\, \overset{\leftrightarrow}{D}_\nu\psi\Big\}\, ,
\qquad 
\partial_\mu \partial_\nu \Big\{\psibar \psi\Big\}
\; ,
\end{align}
on their sextet and triplet components. Terms proportional to the second and third fields
in (\ref{eq:O1}) do not contribute to matrix elements between initial and final states with
the same four-momentum. Since we restrict our analysis to those
cases, we can discard them.
The improved sextet and triplet fields are then obtained by replacing
\be
{T}^{F,\{i\}}_{\mu\nu} \longrightarrow  {T}^{F,\{i\}}_{\mu\nu} + a \delta {T}^{F,\{i\}}_{\mu\nu}\; , \quad i=6,3 
\ee
in Eqs.~(\ref{eq:TF6}) and (\ref{eq:TF3}) respectively, with (no summation over $\mu,\nu$ in (\ref{eq:delta3}))
\bea
\delta {T}^{F,\{6\}}_{\mu\nu} & = & c^{\{6\}}_F(g^2_0)\, {1 \over 8}\, (1-\delta_{\mu\nu}) \psibar\Big[\sigma_{\mu\rho}{F}_{\nu\rho}
+ \sigma_{\nu\rho}{F}_{\mu\rho} \Big]\psi\; , \label{eq:delta6}\\[0.25cm]
\delta {T}^{F,\{3\}}_{\mu\nu} & = & c^{\{3\}}_F(g^2_0) {1 \over 4}
\psibar\Big[\sigma_{\mu\rho}{F}_{\mu\rho} - \sigma_{\nu\rho}{F}_{\nu\rho}\Big]\psi\; . \label{eq:delta3}
\eea
By performing a classical expansion of ${T}^{F,\{i\}}_{\mu\nu}$ in the lattice
spacing~\cite{Luscher:1984xn}, it turns out that the tree-level values of
the coefficients $c^{\{i\}}_F$ are null as well as those multiplying the
other counteterms in Eq.~(\ref{eq:O1}).

\subsection{Mass-degenerate quarks\label{Oa_mass_ndeg}}
When all quarks have the same non-vanishing mass, the renormalization of the coupling
and of the mass can be obtained from Eqs.~(\ref{eq:gmR}) by replacing
\be\label{eq:gmimp}
g^2_0 \longrightarrow \widetilde{g}^2_0 = g_0^2\big(1+b_{g}(g^2_0)\, am_{q}\big)\; , \qquad
m_{q} \longrightarrow {\widetilde m}_{q} = m_{q}\big(1+b_{m}(g^2_0)\, am_{q}\big)\; ,
\ee
where $m_q$ is the subtracted bare mass common to all flavours. The two improvement
coefficients $b_{g}$ and $b_{m}$ must be chosen so to have a mass-independent
renormalization scheme where the renormalized coupling and the mass are free
from O($a$)-effects~\cite{Luscher:1996sc}. Defined this way, both coefficients
do not depend on the renormalization conditions chosen to set $Z_g$ and $Z_m$. The
perturbative expansion of $b_{g}$ starts at O($g_0^2$) since it arises from sea-quark
loop contributions to a gluonic quantity \cite{Sint:1995ch}.

To improve the sextet and triplet parts of $T_{\mu\nu}$, two more O($a$)-counterterms, made of the original gluon and fermion components
multiplied by the quark mass, have to be taken into account.
As a result, the O($a$)-improved fields
read\footnote{We use the same symbol for the unimproved
and the improved energy-momentum tensor field since any ambiguity is resolved from the context.} 
\bea\label{eq:TImunu}
T^{R,\{i\}}_{\mu\nu} & = & Z^{\{i\}}_G(\widetilde{g}^2_0) \big(1 + b_G^{\{i\}}(g^2_0)am_{q}\big) T^{G,\{i\}}_{\mu\nu}\nonumber\\[0.25cm]
& + & Z^{\{i\}}_F(\widetilde{g}^2_0)\, \big(1 +\!b_F^{\{i\}}(g^2_0)am_{q}\big)
\Big\{T^{F,\{i\}}_{\mu\nu} + a \delta T^{F,\{i\}}_{\mu\nu} \Big\}\;, \qquad i=6,3\; .  
\eea
Notice that the renormalization constants appearing in Eq.~(\ref{eq:TImunu}) must be evaluated at the
value $\widetilde{g}^2_0$. On the other hand, it is consistent to evaluate $c^{\{i\}}_F$ and the two
extra coefficients $b_G^{\{i\}}$ and $b_F^{\{i\}}$ either
at $\widetilde{g}^2_0$ or $g^2_0$. The term proportional to $m_{q}\,\delta {T}^{F,\{i\}}_{\mu\nu}$ in
Eq.~(\ref{eq:TImunu}) can be neglected since is O($a^2$). The coefficients $b_G^{\{i\}}$
are null at tree-level because they arise from sea-quark loop contributions.

\subsection{Mass non-degenerate quarks}\label{Oa_mass_nondeg}
When all quark masses are different, the pattern of improvement
is further complicated. An O($a$)-improved
renormalized coupling can be defined from Eqs.~(\ref{eq:gmR})
by replacing~\cite{Bhattacharya:2005rb}
\be
g^2_0 \longrightarrow \widetilde{g}^2_0 = g_0^2\bigg(1+ {b_{g}(g^2_0)\over\Nf}\tr\{a M_q\} \bigg)\; ,
\qquad M_{q}=M_0-m_c(g^2_0) 1\!\!1\; ,
\ee
while the expressions for the renormalized improved quark masses are quite involved and, 
since they are not needed in the following, we refer the interested reader directly to
Eq.~(26) of Ref.~\cite{Bhattacharya:2005rb}. The O($a$)-improved fields turn out to be
\bea
\hspace{-0.5cm} T^{R,\{i\}}_{\mu\nu}\!\!\!\!\! &=&\!\!\!\!\! Z^{\{i\}}_G(\widetilde{g}^2_0)
\Big(1 + {b_G^{\{i\}}(g^2_0)\over N_f} \tr\{a M_{q}\}\Big)
T^{G,\{i\}}_{\mu\nu}\! +\! Z^{\{i\}}_F (\widetilde{g}^2_0) \Bigg[\hat b_F^{\{i\}}(g^2_0)
\sum_{f,g=1}^{N_f} a M_{q}^{fg} T^{F,\{i\}}_{\mu\nu;gf}\nonumber\\
& + &\!\!\!\!\! \Big(1+ {\overline{b}_F^{\{i\}}(g^2_0)\over N_f}\tr\{a M_{q}\}\Big)
\Big\{ T^{F,\{i\}}_{\mu\nu}+a \delta T^{F,\{i\}}_{\mu\nu} \Big\}\Bigg]\, ,\quad i=6,3\; , 
\label{eq:TI_general}
\eea
where $T^{F,\{i\}}_{\mu\nu;fg}$ indicates the flavour non-singlet analogous of $T^{F,\{i\}}_{\mu\nu}$, e.g. 
($f,g=1,\ldots,\Nf$)
\be
T^{F,\{6\}}_{\mu\nu;fg} = (1-\delta_{\mu\nu})
{1 \over 8}\left\{\psibar^f\dirac\mu\big[\nabdbarstar\nu+\nabdbar\nu\big]\psi^g +
                  \psibar^f\dirac\nu\big[\nabdbarstar\mu+\nabdbar\mu\big]\psi^g \right\}\; , 
\ee
and analogously for $T^{F,\{3\}}_{\mu\nu;fg}$. In the mass-degenerate limit $M_{q}=m_{q}\mathbb{1}$,
the expressions of the previous subsection are recovered, up to O($a^2 m$) effects, provided one sets 
$b_F^{\{i\}} = \hat b_F^{\{i\}} + \overline{b}_F^{\{i\}}$. In perturbation theory
the coefficients $\overline{b}_F^{\{i\}}$ are null up to 1-loop order since they originate from sea-quark
loop contributions to fermionic quantities.

\subsection{Improvement conditions\label{sec:imprnp}}
As we have seen in the previous subsections, improving the energy-momentum tensor may 
require the tuning of several parameters. A decoupling of the equations that fix
them, however, occurs naturally within our strategy. Indeed, in the massless limit
chiral symmetry is expected to be either exact or effectively restored when the
temperature $T$ is much larger than the typical scale of the strong interactions,
a few hundred MeV or so. As a consequence, the expectation values of the chiral
non-singlet counterterms in Eq.~(\ref{eq:O1}) either vanish or become quickly
negligible at high temperature. The origin of this result can be traced back to
the more general fact that the thermal theory with massless quarks enjoys
de-facto {\it automatic O($a$)-improvement} at high temperature,
see appendix~\ref{app:autOa} for a detailed discussion in the presence of a
generic number of flavours.

For the particular case we are concerned here, namely the expectation values of the energy-momentum
tensor, stronger results can be
proven\footnote{We thank Martin L\"uscher for suggesting to us this line of argumentation.}.
In the Symanzik effective continuum theory with two or more massless
flavours, it holds
\be\label{eq:WISUP}
\langle \delta {T}^{F,\{i\}}_{\mu\nu} \rangle_{\bsxi,\theta_0} = -{N_f \over N_f^2-1}
\int_{\partial R} d \sigma_{k}
\langle A^a_k(x)  \delta {T}^{a, F,\{i\}}_{\mu\nu} (0) \rangle_{\bsxi,\theta_0}\; , 
\ee
where $\partial R$ is the union of the top and bottom lids of an hyper-cylinder $R$
containing the origin,
$A^a_k=\bar\psi\gamma_k\gamma_5 T^a \psi$ ($k=1,2,3$) with $T^a$ ($a=1,\dots,N_f^2-1$)
being the generators of the group SU($N_f$), and $\delta {T}^{a, F,\{i\}}_{\mu\nu}$
is defined as $\delta {T}^{F,\{i\}}_{\mu\nu}$ but with the replacement
$\psi\rightarrow \gamma_5 T^a\psi$. At large temperature, where the theory has a mass gap,
the integrand in Eq.~(\ref{eq:WISUP}) decreases exponentially
with the distance $|x|$. Its integral is therefore null exactly since the lids can be sent
to infinity, and the conclusions of appendix \ref{app:autOa} are recovered.  
At smaller temperature, where the theory
develops Goldstone bosons~\cite{Donoghue:1991qv}, the integrand decreases power-like in $|x|$.
At large distances, the leading
behaviour is dictated by the single Goldstone-boson contribution so that 
\be\label{eq:corrAT}
\langle A^a_\rho(x)  \delta {T}^{a, F,\{i\}}_{\mu\nu} (0) \rangle_{\bsxi,\theta_0} \propto
\partial_\rho \Big(\partial_\mu \partial_\nu - {1\over 4} \delta_{\mu\nu} \Box  \Big)\Delta(x) + \dots\;  ,
\ee
where $\Delta(x)$ is the free propagator of a massless boson while the dots indicate sub-leading
corrections. At distances $|x|$ much larger than the inverse temperature, 
$\Delta(x)\propto 1/|x|$,  the correlator (\ref{eq:corrAT}) decreases as $|x|^{-4}$, and
the surface integral in (\ref{eq:WISUP}) is again null.

The final important outcome of this analysis is that, if the lattice action
is O($a$)-improved and quarks are massless, the expectation values $\langle {T}^{F,\{i\}}_{\mu\nu} \rangle_{\bsxi,\theta_0}$ are
O($a$)-improved at any temperature as well as the renormalization constants fixed by imposing
Eqs.~(\ref{eq:LatticeWI1}), (\ref{eq:LatticeWI3}) or (\ref{eq:LatticeWI4}).

The improvement coefficients in Eq.~(\ref{eq:TI_general}) can be determined non-perturbatively
by imposing the very same equations (\ref{eq:LatticeWI1}), (\ref{eq:LatticeWI3}) or (\ref{eq:LatticeWI4})
to be valid up to O($a$) terms for several values of the quark masses, and by remembering that the free-energy
density is already improved once the Sheikholeslami-Wohlert term has been included in the action.
A detailed implementation of this strategy to 1-loop order in perturbation theory is discussed in
section~\ref{sec:ImprCoeff}.

\section{Perturbative analysis\label{sec:PT}}
In order to verify analytically the validity of the strategy proposed in this paper, we have
computed the free-energy density and the expectation values of the energy-momentum
tensor components to 1-loop order in lattice perturbation theory in the presence of shifted and
twisted boundary conditions in the infinite spatial volume limit.
The calculation has been carried out by regularizing
gluons with the Wilson plaquette action and quarks with the $O(a)$-improved
Wilson operator, see appendices \ref{App:lat} and \ref{eq:appD} for the definitions
of the actions, free propagators, and lattice vertices. This computation
serves also to determine, for the first time, the renormalization constants of the
sextet and triplet components of $T_{\mu\nu}$ in the O($a$)-improved theory to 1-loop order in perturbation theory,
as well as their $O(a)$-improvement coefficients.
As a byproduct we have confirmed the 1-loop expressions of the
renormalization constants in the unimproved theory which were determined
in Refs.~\cite{Caracciolo:1991cp,Burgio:1996ji,Capitani:1994qn,Capitani:2002mp,Yang:2016xsb}.
Finally the combination of the results in this section with Eqs.~(\ref{eq:LatticeWI1}),
(\ref{eq:LatticeWI3}) or (\ref{eq:LatticeWI4}) leads to perturbatively improved versions
of these non-perturbative renormalization conditions. As a consequence, the numerical
non-perturbative determinations of the renormalization constants are free from 
discretization effects up to order $g_0^2$.

\subsection{Free-energy density}
The 1-loop expansion of the bare free-energy density defined in
Eq.~(\ref{eq:free-energy}) is
\be\label{eq:pt}
f(L_0,\bsxi,\theta) = f^{(0)} +  g_0^2 f^{(1)}\; , 
\ee
where
\be
f^{(0)} =  (N_c^2-1) f^{G (0)} +  N_c N_f f^{F (0)}\; ,\label{eq:ftree}
\ee
and 
\be
f^{(1)} =  (N_c^2-1) \Big[ N_c f^{G(1,N_c)} + {1 \over N_c} f^{G(1,{1 \over N_c})}
                       + N_f f^{F(1,N_f)} \Big]\label{eq:floop}
\ee
are the tree-level and 1-loop contributions respectively\footnote{Throughout this section the arguments
of the various tree-level and 1-loop contributions such as $\vec\xi$, $m_0$, etc. are omitted for
better readability.}. The functions
$f^{G (0)}$ and $f^{F (0)}$ are the tree-level gluonic and fermionic contributions,
$f^{G(1,N_c)}$ and $f^{G(1,{1 \over N_c})}$ are the 1-loop gluonic parts, and
$f^{F(1,N_f)}$ collects the 1-loop fermionic contributions. All these functions are reported in
appendix~\ref{app:freeE}.
In the perturbative computations
presented in this paper we assume always to have $N_f$ quarks with equal masses and twist angles.
The formulas for the generic case, however, can be easily obtained by summing the contributions of
each individual flavour rather than multiplying the single-fermion contribution by
$N_f$, e.g for the free-energy density the terms $N_f f^{F(0)}$ and $N_f f^{F(1,N_f)}$ must be replaced
by the sums over the flavours of the $f^{F(0)}$  and $f^{F(1,N_f)}$ functions  computed
for the mass and the twist angles of each single flavour respectively.

Once the bare free-energy density has been calculated, the 1-loop renormalized expression is
obtained by re-writing the bare parameters $g_0$ and the common bare quark mass $m_{0}$ through the
renormalized counterparts defined in Eq.~(\ref{eq:gmR}). By properly combining Eq.~(\ref{eq:free-energy}),
the renormalized 1-loop expression at finite lattice spacing and its continuum limit, a 1-loop perturbative
improved definition of the free-energy density can also be obtained.

\subsection{Energy-momentum tensor}
The bare expectation values of the sextet and triplet components of $T_{\mu\nu}$ are 
\be
\langle T_{\mu\nu}^{\{i\}}\rangle_{\bsxi,\theta} =   
{\cal T}^{\{i\}(0)}_{\mu\nu} + g_0^2\, {\cal T}^{\{i\}(1)}_{\mu\nu}\; , \quad i=6,3\; , \label{eq:Tsextet}
\ee
where the tree-level values are 
\be
{\cal T}^{\{i\}(0)}_{\mu\nu} =  (N_c^2-1)  {\cal T}^{G\{i\}(0)}_{\mu\nu} + N_c N_f {\cal T}^{F\{i\}(0)}_{\mu\nu}\;,\label{eq:Tfsextet}
\ee
while the 1-loop contributions are
\be
{\cal T}^{\{i\}(1)}_{\mu\nu} =   (N_c^2-1) \Big[ N_c {\cal T}^{G\{i\}(1,N_c)}_{\mu\nu}\! +
  {1 \over N_c} {\cal T}^{G\{i\}(1,{1 \over N_c})}_{\mu\nu}\!
  +\!   N_f {\cal T}^{G\{i\}(1,N_f)}_{\mu\nu}\! +\! N_f {\cal T}^{F\{i\}(1,N_f)}_{\mu\nu}\Big]\; . 
\label{eq:T1lsextet}
\ee
All functions on the r.h.s of Eqs.~(\ref{eq:Tfsextet})--(\ref{eq:T1lsextet}) are given in appendices
\ref{app:freeT6} and \ref{app:freeT3}.

Once the bare parameters of the theory have been renormalized, the definition of the sextet and the triplet
components of $T_{\mu\nu}$ require the calculation of the renormalization constants defined in
Eq.~(\ref{eq:SextetEMT}). At one loop they can be written as 
\begin{equation}\label{ZG_PT}
Z^{\{i\}}_{G} = Z_{G}^{{\{i\}}(0)}+g_0^2\, Z_{G}^{{\{i\}}(1)}\;, \quad
Z^{\{i\}}_{F} = Z_{F}^{{\{i\}}(0)}+g_0^2\; Z_{F}^{{\{i\}}(1)}\; , \quad i=6,3\; ,
\end{equation}
where we can define
\bea
Z^{\{i\}(1)}_{G} & = & N_c Z_{G}^{{\{i\}}(1,N_c)} + {1\over N_c} Z_{G}^{{\{i\}}(1,{1\over N_c})} +
N_f Z_{G}^{{\{i\}}(1,N_f)}\; ,\\[0.25cm]
Z^{\{i\}(1)}_{F} & = &  {(N_c^2-1)\over N_c} Z_{F}^{{\{i\}}(1,N_c)}\; .
\eea
To impose the renormalization conditions in Eqs.~(\ref{eq:LatticeWI1}), (\ref{eq:LatticeWI3}) or
(\ref{eq:LatticeWI4}) in the massless limit
$m_R=0$, we remind that at 1-loop the critical mass is
\be\label{eq:mc1loop}
m_{c} = m^{(0)}_{c} + m^{(1)}_{c} g_0^2  
\ee
where $m^{(0)}_{c}=0$ and~\cite{Panagopoulos:2001fn}
\ba
m^{(1)}_{c} & = & {(N_c^2-1)\over N_c} m^{(1,N_c)}_{c}\; , \\
a m^{(1,N_c)}_{c} & = & -0.16285705871085(1) +c_{\rm sw}\; 0.04348303388205(10)\nonumber\\
& + & c_{\rm sw}^2 \; 0.01809576878142(1)\; .  
\ea
At the order we work, $c_{\rm sw}=0$ or $1$ for the unimproved and improved theory respectively.
The chiral limit in the expressions (\ref{eq:pt})--(\ref{eq:T1lsextet}) is then reached by
requiring that $m_0=m^{(1)}_{c} g_0^2$. In practice, to 1-loop order, that corresponds to Taylor
expand in the bare quark mass the tree-level expressions of the observables, and then to
fix $m_0=m_c^{(0)}=0$ and to replace
\ba
f^{F(1,N_f)} & \longrightarrow & f^{F(1,N_f)} + {\partial f^{F (0)} \over \partial m_0}\, m_{c}^{(1,N_c)}\; ,    \\[0.25cm]
{\cal T}^{F\{i\}(1,N_f)}_{\mu\nu} & \longrightarrow &
{\cal T}^{F\{i\}(1,N_f)}_{\mu\nu} + {\partial {\cal T}^{F\{i\}(0)}_{\mu\nu} \over \partial m_0} \, m_{c}^{(1,N_c)}\; ,\quad i=6,3\; ,
%
\ea
where the derivatives on the r.h.s. of these equations can be found in Eqs.~(\ref{eq:DF0}), (\ref{eq:DT6})
and (\ref{eq:DT3}) respectively.

\subsubsection{Renormalization constants of the sextet}
By imposing Eq.~(\ref{eq:LatticeWI1}) for two different values\footnote{In perturbation
theory we always set $\vec\theta=\vec 0$ since we work in the infinite spatial volume limit.} of
$\theta_0=\theta^A_0,\theta^B_0$, the tree-level values can be defined as 
\begin{equation}\label{eq:free_sext}
  Z_G^{\{6\}(0)} = -{1 \over {\cal T}^{G\{6\}(0)}_{0k}}\; {\Delta f^{G (0)} \over \Delta \xi_k}
  \qquad \mbox{and} \qquad
  Z_F^{\{6\}(0)} = -{1 \over {\cal T}^{F\{6\}(0)}_{0k}}\; {\Delta f^{F (0)} \over \Delta \xi_k} 
\end{equation}
where the discrete derivative $\Delta$ with respect to the shift is defined as
in Eq.~(\ref{eq:LatticeXiDerivative}). As expected, in the limit
$L_0/a\rightarrow\infty$ it holds $Z_{G}^{\{6\}(0)}=Z_{F}^{\{6\}(0)}=1$. The pure gluonic
1-loop contributions are
\be
Z_{G}^{\{6\}(1,N_c)} = 
-{1 \over {\cal T}^{G\{6\}(0)}_{0k}}\; \Big\{Z_G^{\{6\}(0)}\, {\cal T}_{0k}^{G\{6\}(1,N_c)} + {\Delta f^{G(1,N_c)} \over \Delta \xi_k}\Big\}\; , \\
\ee
and $Z_{G}^{\{6\}(1,{1\over N_c})}$ which has the very same expression as $Z_{G}^{\{6\}(1,N_c)}$ once $(1,N_c)\rightarrow (1,{1\over N_c})$.

The last two terms, $ Z_{G}^{\{6\}(1,N_f)} $ and $Z_{F}^{\{6\}(1,N_c)}$, depend on the interaction
between quarks and gluons. If we define the combination
\be
d^{\{6\}} = - \Big\{ Z_G^{\{6\}(0)} {\cal T}_{0k}^{G\{6\}(1,N_f)} + Z_F^{\{6\}(0)} {\cal T}_{0k}^{F\{6\}(1,N_f)}  + {\Delta f^{F(1,N_f)} \over \Delta \xi_k}\Big\}\; , 
\ee
we obtain
\be\label{eq:Zg61loop}
Z_{G}^{\{6\}(1,N_f)} = {1\over {\cal T}^{G\{6\}(0)}_{0k}}\Big\{d^{\{6\}} - Z_{F}^{\{6\}(1,N_c)} {\cal T}_{0k}^{F\{6\}(0)}\Big\}\; , 
\ee
where the renormalization constant of the fermion component is
\be\label{eq:Zf1loop}
Z_{F}^{\{6\}(1,N_c)}={ d^{\{6\}}(\theta^A_0) -  d^{\{6\}}(\theta^B_0) \over {\cal T}_{0k}^{F\{6\}(0)}(\theta^A_0) - {\cal T}_{0k}^{F\{6\}(0)}(\theta^B_0) }\; , 
\ee
and in the last equation we have explicitly indicated the dependence on $\theta_0$ since is
the only one where two different values are needed.

The values of the various terms which define $Z^{\{6\}}_{G}$ and $Z^{\{6\}}_{F}$ have been computed numerically on lattices with temporal
extension $L_0/a$ ranging from $4$ to $32$ in steps of $2$ and spatial size  $L_1=L_2=L_3= R L_0$ with
$R=6$, $8$, $10$ and $12$. The calculations have been carried out for two values of the shift,
$\vec\xi = (1,0,0)$ and $(1/2,1/2,0)$,  and for three values of the fermionic phase in the temporal
direction, $\theta_0=0$, $\pi/16$ and $\pi/4$. At fixed value of the shift, data for the three values of
$\theta_0$ have been analyzed by considering two independent differences. This large amount of data
allowed us to extrapolate the results to infinite spatial volume and to $a/L_0\rightarrow 0$ limit with
confidence. The latter extrapolation has been performed by fitting the results in powers of $(a/L_0)^2$
supplemented by terms multiplied by $\ln{(a/L_0)}$ for the 1-loop coefficients. The required lattice sums
have been computed in coordinate space~\cite{Luscher:1995zz} after having used the Fast Fourier
Transform algorithm for computing the gluon and quark propagators.
A hierarchical procedure for sums has been implemented in order to preserve a high numerical
accuracy. For some of them, however,  we needed to run in quadruple precision due to large cancellations
taking place at large volumes.

The final results for the various contributions to $Z^{\{6\}}_{G}$ and $Z^{\{6\}}_{F}$ in the limit
$a/L_0\rightarrow 0$ are listed in
Table \ref{tab:tripsext} together with the analogous values present in the
literature~\cite{Caracciolo:1991cp,Burgio:1996ji,Capitani:1994qn,Capitani:2002mp,Yang:2016xsb}.
Our error bars have been estimated by changing the fit range in $(a/L_0)$, by considering the spread over
the two differences in $\theta_0$, and by analyzing the dependence on $R$. For all values that
can be compared with Refs.~\cite{Caracciolo:1991cp,Burgio:1996ji,Capitani:1994qn,Capitani:2002mp,Yang:2016xsb}, the agreement is excellent.

\begin{table}[t!]
\centering
\begin{tabular}{|c|cc|cc|}
  \hline
   & \multicolumn{2}{|c|}{Sextet ($i=6$)} & \multicolumn{2}{|c|}{Triplet ($i=3$)}\\ 
  & This work & Ref~\cite{Caracciolo:1991cp,Burgio:1996ji,Capitani:1994qn,Capitani:2002mp,Yang:2016xsb} &
  This work& Ref~\cite{Caracciolo:1991cp,Burgio:1996ji,Capitani:1994qn,Capitani:2002mp,Yang:2016xsb} \\ 
  \hline
  $Z_G^{\{i\}(0)}$               & 1.000000(5)   & 1            & 1.000000(5) & 1             \\[0.25cm] 
  $Z_{G}^{\{i\}(1,N_c)}$          & 0.10414(3)    & 0.10413887   & 0.09773(3)  & 0.09772334 \\[0.25cm]
  $Z_{G}^{\{i\}(1,{1\over N_c})}$   & -0.125000(1)  & -1/8         & -0.157495(1) & -0.15749516  \\[0.25cm]
  $Z_{G,c_{\rm sw}=0}^{\{i\}(1,N_f)}$ & 0.010827(1)   & 0.01082699  & 0.00601(2)  & 0.006010835  \\[0.25cm]
  $Z_{G,c_{\rm sw}=1}^{\{i\}(1,N_f)}$ & 0.0301785(25) & ---        & 0.02466(2)  & ---          \\[0.25cm]
  \hline
  $Z_F^{\{i\}(0)}$               & 1.000000(5)   & 1          & 1.000000(5) &       1       \\[0.25cm] 
  $Z_{F,c_{\rm sw}=0}^{\{i\}(1,N_c)}$ & -0.01474(1)   & -0.01473   & -0.03167(3) &    -0.03169   \\[0.25cm] 
  $Z_{F,c_{\rm sw}=1}^{\{i\}(1,N_c)}$ & 0.005282(4)   & ---        & -0.0109(1)  &      ---     \\[0.25cm] 
  \hline
\end{tabular}
\caption{Numerical values of the coefficients of the perturbative expressions
  of $Z^{\{i\}}_G$ and $Z^{\{i\}}_F$ computed in this work for $c_{\rm sw}=0$ and $1$ corresponding to the unimproved and
  improved theory respectively. Results from
  Refs.~\cite{Caracciolo:1991cp,Burgio:1996ji,Capitani:1994qn,Capitani:2002mp,Yang:2016xsb} are also shown for comparison.} 
\label{tab:tripsext}
\end{table}

\subsubsection{Renormalization constants of the the triplet}
By combining Eqs.~(\ref{eq:LatticeWI1}) and (\ref{eq:LatticeWI3}) or (\ref{eq:LatticeWI4}), the renormalization constants
of the triplet can be determined analogously to the previous subsection. The Eqs.~(\ref{eq:free_sext})--(\ref{eq:Zf1loop})
hold provided the representation index changes, $6\rightarrow 3$, and the
various components of ${\cal T}_{0k}$ are multiplied by $\xi_k/(1-\xi_k^2)$
or $\xi_k$ when Eq.~(\ref{eq:LatticeWI3}) or (\ref{eq:LatticeWI4}) are used respectively.
A different option is, however, possible. Once the renormalization constants of the
sextet have been computed, the ones of the triplet can be determined directly imposing
Eq.~(\ref{eq:LatticeWI3}) or (\ref{eq:LatticeWI4}) for two different values of $\theta_0=\theta^A_0,\theta^B_0$.
In this first case the tree-level, values are 
\begin{equation}
  Z_G^{\{3\}(0)} =  Z_G^{\{6\}(0)} {1-\xi_k^2 \over \xi_k}\, {{\cal T}^{G\{6\}(0)}_{0k} \over {\cal T}^{G\{3\}(0)}_{0k}}\label{eq:zg0}\; , 
  \qquad
  Z_F^{\{3\}(0)} = Z_F^{\{6\}(0)} {1-\xi_k^2\over \xi_k}\, {{\cal T}^{F\{6\}(0)}_{0k} \over {\cal T}^{F\{3\}(0)}_{0k}}\; , 
\end{equation}
with $Z_{G}^{\{3\}(0)}=Z_{F}^{\{3\}(0)}=1$ in the limit $a/L_0 \rightarrow 0$. The pure gluonic contributions are
\be
\hspace{-0.4cm} Z_{G}^{\{3\}(1,N_c)}\!\!\!\! = \!\! Z_{G}^{\{6\}(1,N_c)}\! {Z_G^{\{3\}(0)} \over Z_G^{\{6\}(0)}}  + {1\over {\cal T}_{0k}^{G\{3\}(0)}}\!
\Big\{\!{1\!\!-\!\!\xi_k^2\over\xi_k}\, Z_G^{\{6\}(0)} {\cal T}_{0k}^{G\{6\}(1,N_c)}\!\!\!  -
Z_G^{\{3\}(0)} {\cal T}_{0k}^{G\{3\}(1,N_c)}\!\Big\}\!,
\ee
and $Z_{G}^{\{3\}(1,{1\over N_c})}$ which has the very same expression as $Z_{G}^{\{3\}(1,N_c)}$ once
$(1,N_c)\rightarrow (1,{1\over N_c})$.
The last two terms, $Z_{G}^{\{3\}(1,N_f)}$ and $Z_{F}^{\{3\}(1,N_c)}$, depend on the interaction
between quarks and gluons. If we define the combination
\bea
\hspace{-0.375cm} d^{\{3\}} & = & Z_G^{\{6\}(0)} {\cal T}_{0k}^{G\{6\}(1,N_f)} + Z_F^{\{6\}(0)} {\cal T}_{0k}^{F\{6\}(1,N_f)}  +
               Z_G^{\{6\}(1,N_f)} {\cal T}_{0k}^{G\{6\}(0)}\nonumber\\[0.25cm]
& + &\!\!\!\!\! Z_F^{\{6\}(1,N_c)} {\cal T}_{0k}^{F\{6\}(0)}\!\! -\!
{\xi_k\over 1-\xi_k^2}\,\Big\{Z_G^{\{3\}(0)} {\cal T}_{0k}^{G\{3\}(1,N_f)}\! + Z_F^{\{3\}(0)} {\cal T}_{0k}^{F\{3\}(1,N_f)} \Big\}\, , 
\eea
we obtain
\be\label{eq:Zg31loop}
Z_{G}^{\{3\}(1,N_f)} ={1 \over {\cal T}^{G\{3\}(0)}_{0k}}  \Big\{{1-\xi_k^2 \over \xi_k}\, d^{\{3\}} -
Z_{F}^{\{3\}(1,N_c)}  {\cal T}_{0k}^{F\{3\}(0)}   \Big\}\; , 
\ee
where the renormalization constant of the fermion component is
\be
Z_{F}^{\{3\}(1,N_c)}= {1-\xi_k^2 \over \xi_k}\, { d^{\{3\}}(\theta^A_0) -  d^{\{3\}}(\theta^B_0) \over
{\cal T}_{0k}^{F\{3\}(0)}(\theta^A_0) - {\cal T}_{0k}^{F\{3\}(0)}(\theta^B_0) }\; , \label{eq:zf1}
\ee
and in the last equation we have explicitly indicated the dependence on the set of twisted angles.
It is important to notice that the values of the shift $\vec\xi$, and of the angles $\theta^A_0$
are $\theta^B_0$ used for fixing the sextet and the triplet renormalization constants can be in general
different. If the renormalization condition (\ref{eq:LatticeWI4}) is used instead of (\ref{eq:LatticeWI3}), the
Eqs.~(\ref{eq:zg0})--(\ref{eq:zf1}) remain valid but with the replacements
$\xi_k/(1-\xi_k^2) \rightarrow \xi_k$ and $k \rightarrow j$ in the second subscript index of the
triplet components.

The numerical value of the various contributions to $Z^{\{3\}}_G$ and $Z^{\{3\}}_F$ have been computed
on lattices with temporal extension $L_0/a$ ranging from $4$ to $32$ in steps of $2$ and spatial size
$L_1=L_2=L_3= R L_0$ with $R=$ $10$ and $15$. The calculations have been carried out for
$\vec\xi = (2,0,0)$ so to satisfy the constraint in Eq.~(\ref{eq:PeriodicCondition}). Three values of
the fermionic phase, $\theta_0=0$, $\pi/16$ and $\pi/4$, have been considered and
a non vanishing phase along the direction $\hat 1$ has been chosen according to the third
constraint in Eq.~(\ref{eq:allb1}). Analogously to the sextet case, data for
three values of $\theta_0$ have been analyzed by considering two possible independent differences.
The results show no relevant dependence from the spatial volume. The $a/L_0\rightarrow 0$ extrapolation
has been performed again by fitting the results in powers of $(a/L_0)^2$ supplemented by terms multiplied
by $\ln{(a/L_0)}$ for the 1-loop coefficients. As for the sextet, the lattice sums have been computed in
coordinate space, and also in this case a hierarchical procedure was implemented. 

The final results for the various contributions to $Z^{\{3\}}_G$ and $Z^{\{3\}}_F$ in the limit
$a/L_0\rightarrow 0$ are listed in
Table~\ref{tab:tripsext} together with the analogous ones present in the
literature~\cite{Caracciolo:1991cp,Burgio:1996ji,Capitani:1994qn,Capitani:2002mp,Yang:2016xsb}. The error
bars have been estimated by changing the fit range in $(a/L_0)$, and by considering the spread over
the two differences in $\theta_0$. Whenever a comparison with Refs.~\cite{Caracciolo:1991cp,Burgio:1996ji,Capitani:1994qn,Capitani:2002mp,Yang:2016xsb} is
possible, the agreement is excellent.

\subsubsection{Improvement coefficients \label{sec:ImprCoeff}}
We conclude the perturbative analysis of the strategy proposed in this paper by
computing to 1-loop order the improvement coefficients of the sextet and triplet
components of $T_{\mu\nu}$ introduced in Eq.~(\ref{eq:TI_general}). The terms
$\delta {\mathcal{T}}^{F,\{i\}}_{\mu\nu}$ do not contribute to this order,
since the coefficients $c^{\{i\}}_F(g^2_0)$, as mentioned in
section~\ref{Oa_massless}, are null at tree level as well as, for our
choice of boundary conditions, the associated fields.
The two $\overline{b}_F^{\{i\}}(g^2_0)$ vanish as well. The other coefficients in
Eq.~(\ref{eq:TI_general}), or equivalently Eq.~(\ref{eq:TImunu}), can be written as
\begin{equation}\label{b6impr}
  b_G^{\{i\}} (g^2_0) =  N_f b_G^{\{i\}(1,N_f)} g_0^2\; , \quad
  b_F^{\{i\}} (g^2_0) =  1 + {N_c^2-1\over N_c}\, b_F^{\{i\}(1,N_c)} g_0^2\;, \quad
  i=6,3\; ,  
\end{equation}
where the tree-level values of the $b_F^{\{i\}}$ have been fixed so to remove the
O($a$) terms in the tree-level expressions of the fermion components of
$T_{\mu\nu}$.
They can be computed by imposing Eqs.~(\ref{eq:LatticeWI1}),
(\ref{eq:LatticeWI3}) or (\ref{eq:LatticeWI4}) up to O($a$) terms in the simpler case
where all quarks have the same non-vanishing mass $m_R\neq 0$.

To this aim, by using Eqs.~(\ref{eq:gmR}) and (\ref{eq:gmimp}), we remind that the
O($a$)-improved quark mass is given by
\be\label{eq:mRimp}
m_R = Z_{m}(g^2_0,a \mu)\, \big(1+b_{m}(g^2_0)\, am_{q}\big)\, m_{q}\; ,
\ee
where at 1-loop in perturbation theory~\cite{Sint:1997jx}
\begin{equation}\label{Zmexpand}
  Z_m(g^2_0,a\mu) = 1+{N_c^2-1\over N_c} Z_m^{(1,N_c)} g_0^2\; ,  
\qquad
  Z_m^{(1,N_c)} = -{3\over (4\pi)^2} \ln (a\mu)\; , 
\end{equation}
\begin{equation}\label{bmexpand}
  b_m(g^2_0) = -{1\over 2} + {N_c^2-1\over N_c}   b_m^{(1,N_c)} g_0^2\; , \qquad
 \!\!\!\! b_m^{(1,N_c)}= -0.036085(10)\, .
\end{equation}
Notice that the values of the improvement coefficients do not depend on 
the scale $\mu$. By remembering that
$m_q=m_0-m_{c}(g^2_0)$, with $m_{c}(g^2_0)$ being the critical mass 
in Eq.~(\ref{eq:mc1loop}), one easily finds that the bare mass which solves
Eq.~(\ref{eq:mRimp}) at fixed renormalized quark mass is $m_0(g^2_0)= m_0^{(0)}+m_0^{(1)} g_0^2$ where~\cite{Sint:1997jx}
\be
a m_0^{(0)} = 1- \sqrt{1-2 a m_R }\; ,
\ee
and
\ba
m_0^{(1)} & = & {N_c^2-1\over N_c} m_0^{(1,N_c)}\; , \\
a m_0^{(1,N_c)}\!\! & = &\!\! a m_{c}^{(1,N_c)}
- {a m_R\, Z_m^{(1,N_c)} + 2 b_m^{(1,N_c)} \Big(1 - a m_R - \sqrt{1-2 a m_R }\Big)
  \over \sqrt{1-2 a m_R } }\; .
\ea
The free-energy density and the expectation values of the energy-momentum tensor
at bare mass $m_0(g_0)$ are obtained by evaluating the expressions (\ref{eq:pt})--(\ref{eq:T1lsextet})
at $m_0=m_0^{(0)}$ and by replacing 
\ba
f^{F(1,N_f)} & \longrightarrow & f^{F(1,N_f)} + {\partial f^{F (0)} \over \partial m_0}\, m_{0}^{(1,N_c)}\; ,    \\[0.25cm]
{\cal T}^{F\{i\}(1,N_f)}_{\mu\nu} & \longrightarrow &
{\cal T}^{F\{i\}(1,N_f)}_{\mu\nu} + {\partial {\cal T}^{F\{i\}(0)}_{\mu\nu} \over \partial m_0} \, m_{0}^{(1,N_c)}\; ,\quad i=6,3\; ,
%
\ea
where the derivatives on the r.h.s. of these equations can be found in Eqs.~(\ref{eq:DF0}), (\ref{eq:DT6})
and (\ref{eq:DT3}) respectively.

The solutions of the Eqs.~(\ref{eq:LatticeWI1}), (\ref{eq:LatticeWI3})
or (\ref{eq:LatticeWI4}) at finite quark mass are obtained by replacing 
\begin{table}[t!]
\centering
\begin{tabular}{|c|c|c|} 
\hline
& Sextet &  Triplet \\
& $i=6$  &  $i=3$   \\ 
\hline
 $b_G^{\{i\}(1,N_f)}$ & -0.012(1) & -0.011(2) \\[0.125cm] 
 $b_F^{\{i\}(1,N_c)}$ &  0.051(2) &  0.056(6) \\[0.125cm] 
\hline
\end{tabular}
\caption{Numerical values of the coefficients $b_G^{\{i\}(1,N_f)}$ and $b_F^{\{i\}(1,N_c)}$ to 1-loop order.}
\label{impr_coeff}
\end{table}
\ba
Z_F^{\{i\}(0)} & \longrightarrow &Z_F^{\{i\}(0)} (1 + a m_0^{(0)})\; ,\\[0.25cm] 
Z_{G}^{\{i\}(1,N_f)} & \longrightarrow & Z_{G}^{\{i\}(1,N_f)}  +  Z_G^{\{i\}(0)}
b_G^{\{i\}(1,N_f)} a m_0^{(0)}\; ,\\[0.25cm]
Z_{F}^{\{i\}(1,N_c)} & \longrightarrow & Z_{F}^{\{i\}(1,N_c)} (1 + a m_0^{(0)})\nonumber\\[0.25cm]
& + & Z_F^{\{i\}(0)} (a m_0^{(1,N_c)} - a m_c^{(1,N_c)} + b_F^{\{i\}(1,N_c)} a m_{0} ^{(0)})\; ,
\ea
in Eqs.~(\ref{eq:free_sext})--(\ref{eq:zf1}).
The improvement coefficients are finally determined by inserting into the solutions 
the values of the renormalization constants obtained in the previous subsections,
and then by solving for $b_G^{\{i\}(1,N_f)}$ and $b_F^{\{i\}(1,N_c)}$.
In Table~\ref{impr_coeff} we list the numerical values of $b_G^{\{i\}(1,N_f)}$ and
$b_F^{\{i\}(1,N_c)}$ that we have obtained in the limit $a/L_0\rightarrow 0$ by numerical
computations and analyses analogous to those carried out in the previous two subsections.

\section{Conclusions\label{sec:concl}}
Shifted boundary conditions in the presence of an imaginary chemical potential offer an
extremely powerful tool to non-perturbatively renormalize composite operators on the lattice.
In this work we have applied this framework to the case of the energy-momentum tensor.
The strategy
proposed here is the natural extension of the one already applied successfully to the $\SUthree$ Yang-Mills theory~\cite{Giusti:2015daa}.
The inclusion of quarks, however, complicates the problem because the gluonic and fermionic parts of the
tensor mix together. Introducing a non-zero imaginary chemical potential gives the
handle to solve that problem since, via a conserved charge, it couples differently to quark and gluons, in particular directly
to quarks but only indirectly to gluons through their interaction with quarks. Ward identities
can thus be written, both for the sextet and the triplet, which are different enough
to resolve the mixing between the gluonic and the fermionic parts so that the computation
becomes feasible non-perturbatively.

In view of that application and in order to check the whole construction, we have applied
the method in lattice perturbation theory by computing the renormalization constants
and the O($a$)-improvement coefficients of the  sextet and triplet components of
the energy-momentum tensor to 1-loop order. The agreement with the results in
the literature for the unimproved theory represents a very non-trivial test
of the entire strategy proposed in this paper. A further confirmation of theory expectations
is the $\vec\xi$ and $\theta$ independence of the renormalization constants once
extrapolated to the $a/L_0\rightarrow 0$ limit. An important byproduct of these computations is
the possibility of defining 1-loop perturbative improved estimators of the renormalization
constants and of the expectation values of the energy-momentum tensor, so to reduce discretization
effects in their non-perturbative determinations.

Once fixed so to satisfy the WIs, the renormalization constants are part of the definition of the energy-momentum
field itself. Up to discretization errors, they do not depend on the particular correlator or kinematic conditions they were
employed to fix them. They can be directly used to renormalize the energy-momentum tensor inserted at a
physical distance from other fields in any correlator of QCD, e.g at zero or non-zero temperature with
or without chemical potential.

\section{Acknowledgments}
We thank Martin L\"uscher for many illuminating discussions, especially on topics in sections \ref{sec:renoTc} and
\ref{sec:imprnp}, and for comments on a preliminary version of this paper.
The numerical integrals needed in lattice perturbation theory have been computed on the PC clusters Marconi at CINECA
(CINECA- INFN and CINECA-Bicocca agreements) and Wilson at Milano-Bicocca. We thank these institutions for the computer
resources and the technical support. We also acknowledge PRACE and ISCRA for awarding us access to MareNostrum at Barcelona
Supercomputing Center (BSC), Spain (n. 2018194651) and to Marconi at CINECA (EoSQCD) respectively, where the
non-perturbative computations are being performed. We acknowledge partial support by the INFN
project ``High performance data network''.


\appendix

\section{Conventions and useful identities\label{App:conv}}
Here we summarize the conventions for the generators of the SU($N_c$) group and for the Dirac matrices,
$\gamma_\mu$, together with some standard identities that we have used in the 1-loop perturbative
computation. Let $T_a$, $a=1,\ldots, (N_c^2-1)$, be the hermitean traceless generators of the group
SU($N_c$) normalized as
\begin{equation}
\Tr [ T_a T_b]= {1\over2} \delta_{ab}\; .
\end{equation}
Their commutation and anti-commutation relations are\footnote{Summation over repeated indices is
always understood.}
\begin{equation}
[ T_a, T_b]= i f_{abc} T_c\; , \qquad \qquad \qquad
\{ T_a, T_b\}= {1\over N_c} 1\!\! 1 \delta_{ab}+ d_{abc} T_c\; ,
\end{equation}
where $f_{abc}$ is the completely antisymmetric tensor of the structure constants while 
$d_{abc}$ is completely symmetric. It then holds
\begin{equation}
\Tr [ T_a T_b T_c]= {1\over4} (i f_{abc}+ d_{abc})\; ,
\end{equation}
\begin{equation}
\Tr [ T_a T_a T_b T_b]= {(N_c^2-1)^2\over 4 N_c}\; ,
\qquad
\Tr [ T_a T_b T_a T_b]= -{(N_c^2-1)\over 4 N_c}\; ,
\end{equation}
\begin{equation}
\Tr [ T_a T_b T_c T_d]= 
{1\over4} \Big\{{1\over N_c} \delta_{ab} \delta_{cd} 
+{1\over2} \Big(-f_{abe} f_{cde} +d_{abe} d_{cde} +i f_{abe} d_{cde}+ i d_{abe} f_{cde}
\Big)\Big\}\; .
\end{equation}
Other useful identities are
\begin{equation}
f_{ace} f_{bce}=N_c \delta_{ab}\; , 
\qquad
d_{ace} d_{bce}={N_c^2-4\over N_c} \delta_{ab}\; ,   
\end{equation}
as well as $d_{aae}=0$ and $f_{ace} d_{bce}=0$.\\

By defining $\gamma_5=\gamma_0\gamma_1\gamma_2\gamma_3$, the 
Euclidean anti-commutation relations
\begin{equation}\label{Clifford}
\{ \gamma_\mu,\gamma_\nu \} = 2 \delta_{\mu\nu} 1\!\! 1
\end{equation} 
imply
\begin{equation}
\label{eq:g5}
\{\gamma_5, \gamma_\mu\}= 0\; , \qquad \qquad
\gamma_5^2 = 1\!\! 1\; , 
\qquad \qquad
\gamma_5\gamma_\mu\gamma_5= -\gamma_\mu\; . 
\end{equation}
Useful trace identities are
\begin{equation}
\Tr [\gamma_\mu \gamma_\nu] = 4 \delta_{\mu\nu}\;, \qquad
\Tr [\gamma_\alpha \gamma_\beta \gamma_\delta \gamma_\sigma] = 
4  [ \delta_{\alpha\beta}\delta_{\delta\sigma}-\delta_{\alpha\delta}\delta_{\beta\sigma}+\delta_{\alpha\sigma}\delta_{\beta\delta}]\; ,
\end{equation}
and 
\begin{eqnarray}
\Tr [\gamma_\alpha \gamma_\beta \gamma_\delta \gamma_\rho \gamma_\sigma \gamma_\tau] & = &   
\delta_{\alpha\beta} \Tr[\gamma_\delta \gamma_\rho \gamma_\sigma \gamma_\tau] 
-\delta_{\alpha\delta}\Tr [\gamma_\beta \gamma_\rho \gamma_\sigma \gamma_\tau]
  +\delta_{\alpha\rho} \Tr [\gamma_\beta \gamma_\delta \gamma_\sigma \gamma_\tau] \nonumber \\[0.125cm]
& &\!\!\!\!\! - \delta_{\alpha\sigma} \Tr [\gamma_\beta \gamma_\delta \gamma_\rho \gamma_\tau]
+\delta_{\alpha\tau}\Tr [\gamma_\beta \gamma_\delta \gamma_\rho \gamma_\sigma ]\; .
\end{eqnarray}
while, thanks to Eqs.~(\ref{eq:g5}), traces of products of an odd number of $\gamma$-matrices
vanish.

\section{Continuum theory \label{App:cont}}
In the Euclidean space-time, the path integral of QCD is defined as 
\be\label{eq:Zcont}
Z = \int D A\, D \psibar \, D \psi \, 
D \bar c \, D c\; e^{-S}\; , 
\ee
where the integration measures on the various fields are defined as usual. The 
action is defined as\footnote{Throughout the paper we assume the strong
CP violation term to be absent.}
\be\label{eq:SQCD}
S=\int d^4 x\, {\cal L}(x)\; , \qquad {\cal L} = {\cal L}^G + {\cal L}^{GF} + {\cal L}^{FP} + {\cal L}^{F}\; , 
\ee 
with 
\ba
{\cal L}^G & = & \displaystyle {1 \over 2 g_0^2} \Tr\Big[F_{\mu\nu}\, F_{\mu\nu} \Big]\; , \\[0.25cm]
{\cal L}^{GF} & = & {\lambda_0 \over g^2_0} \Tr\Big[\partial_\mu A_\mu\, \partial_\nu A_\nu\Big]\; ,\\[0.25cm]
{\cal L}^{FP} & = & - {2 \over g_0^2} \Tr\Big[\bar c\, \partial_\mu {\cal D}_\mu c \Big]\;,\\[0.25cm]
{\cal L}^F & = & \psibar\big[\dirac\mu D_\mu + M_0\big]\psi\; ,
\ea
where $g_0$ is the bare coupling constant, $\lambda_0$ is the gauge-fixing 
parameter, the trace is over the color index and  
\ba\label{eq:basics}
F_{\mu\nu}  & = & \partial_\mu A_\nu - 
\partial_\nu A_\mu -i\, [A_\mu,A_\nu]\; , \qquad A_\mu=A_\mu^a\, T^a \\[0.25cm]
D_\mu  & = & \partial_\mu\,  - i\, A_\mu \; , \\[0.25cm]
{\cal D}_\mu c & = & \partial_\mu\, c - i\, [A_\mu,c]\;, \qquad c =c^a  T^a\; .
\ea
The quark and anti-quark fields, $\psi$ and $\psibar$ have  $\Nf$-flavour components
$\psi^f,\psibar^f$, $f=1,\ldots,\Nf$ and, accordingly, the mass matrix
$M_0=\text{diag}(m_{0,1},m_{0,2},m_{0,3},\ldots)$ is a $\Nf\times\Nf$ matrix, whose
entries on the diagonal are the bare quark masses. It turns out to be
useful also to define
\be\label{eq:Dbf}
\overleftrightarrow{D}_\mu = D_\mu - \overleftarrow{D}_\mu\; , \qquad
\overleftarrow{D}_\mu = \overleftarrow{\partial}_\mu + i A_\mu\; . 
\ee

\subsection{Dimensional regularization}
Here we report the essential formulas of dimensional regularization which are needed
in this paper. We follow the conventions of Ref.~\cite{Capitani:1998mq}, see also
Ref.~\cite{Weisz:2010nr} for a recent review. By replacing $\int d^4 x \rightarrow \int d^D x$, one
defines the renormalized coupling $g$ and quark mass matrix $M$ as 
\ba
g_0^2 & = & \mu^{2\epsilon}\, g^2 {\cal Z}^{-1}_g\, ,\label{eq:renDRg}\\
M_0 & = & M {\cal Z}^{-1}_m\; ,,\label{eq:renDRm}
\ea
where $D=4-2\epsilon$ and $\mu$ is a mass parameter. A generic renormalization constant, including
those of composite operators, is expanded in $g^2$. In the MS scheme is then implicitly fixed by
requiring it to be a polynomial in $1/\epsilon$ with no constant term
\be
{\cal Z } = 1 + \sum_{k=1}^{\infty} {\cal Z}^{(k)} g^{2k}\; ,\qquad
{\cal Z}^{(k)} = \sum_{j=1}^{k} {\cal Z}^{(k,j)} {1\over \epsilon^j}\; .   
\ee
The $\beta$-function of the theory is defined as  
\ba\label{eq:betaeps}
\beta(\epsilon, g)  & = & \mu {\partial g \over \partial \mu} = 
-\epsilon g \left\{1 - {g \over 2} {\partial \over \partial g} \ln{Z_g} \right\}^{-1}
=  -\epsilon g + \beta(g)\; ,
\ea
where
\be
\beta(g) = -g^3 \sum_{k=0}^{\infty} b_k g^{2k}\label{eq:betagsol}\;  
\ee
with the first coefficient given by
\be\label{eq:bo1}
b_0 = {1 \over (4\pi)^2}\Big\{{11 \over 3} N_c - {2\over 3} N_f\Big\}\; .  
\ee
The anomalous dimension of the quark mass is defined as
\be\label{eq:gam_m}
\gamma_m = \beta(\epsilon, g) {\partial\over \partial g} \ln{{\cal Z}_m} = - g^2 \sum_{k=0}^{\infty} d_k g^{2k}\; , 
\ee
where the first coefficient is
\ba
d_0  = {3\over (4\pi)^2} {N_c^2-1\over N_c}\;.
\ea

\section{Lattice  theory \label{App:lat}}
The action of the lattice theory reads 
\begin{equation}\label{eq:LatAction}
  S=S^G + S^F\; ,
\end{equation}
where $S^G$ and $S^F$ are the gluonic and fermionic contributions respectively.
For the gluonic one we consider the Wilson plaquette action 
\begin{equation}
\label{eq:SG}
\Sg= {1\over g_0^2} \sum_x \sum_{\mu,\nu} {\rm Re}\,\Tr\big\{1\!\! 1-U_{\mu\nu}(x)\big\}  \;,
\end{equation} 
where $g_0$ is the bare gauge coupling. The plaquette field is defined by 
\begin{equation}
  U_{\mu\nu}(x)=U_\mu(x)U_\nu(x+ a \hat{\mu})U_\mu^\dag(x+ a \hat{\nu})U_\nu^\dag(x)\; ,
\end{equation}
where $\hat{\mu},\hat{\nu}$ are unit vectors oriented along the directions $\mu,\nu$ respectively.
The fermionic part reads
\begin{equation}\label{eq:fermionicAction}
	\Sf=a^4\sum_x \psibar(x) (D+M_0)\psi(x)\; , 
\end{equation}
where $M_0$ is the bare quark mass matrix, and for $D$ we choose the O($a$)-improved
Wilson-Dirac operator 
\begin{equation}
  \label{eq:Dirac}
  D=D_{\rm w} + a D_{\rm sw}\; . 
\end{equation}
The first operator on  the r.h.s is the massless Wilson-Dirac operator defined by
\begin{equation}
  D_{\rm w} =
  {1 \over 2}\big\{\dirac\mu(\nabstar\mu+\nab\mu)-a\nabstar\mu \nab\mu\big\}\; ,
\end{equation}
where $\nabla_\mu^*,\nabla_\mu$ are covariant lattice derivatives acting on the quark fields as follows
\ba
a \nab\mu \psi(x) & = & U_\mu(x)\psi(x+ a \hat{\mu})-\psi(x)\; ,\nonumber \\[0.25cm]
a \nabstar\mu \psi(x) & = & \psi(x) - U^\dag_\mu(x- a \hat{\mu})\psi(x - a \hat{\mu})\; .
\label{eq:fwd-nablas}
\ea
The second term is the Sheikholeslami-Wohlert operator defined by~\cite{Sheikholeslami:1985ij}
\begin{equation}
	\label{eq:DiracSW}
	D_{\rm sw}\psi(x) = c_{\rm sw}(g_0) {1 \over 4}
	\sigma_{\mu\nu} \widehat F_{\mu\nu}(x)\psi(x)\; ,
\end{equation}
where $\sigma_{\mu\nu}=\frac{i}{2}[\dirac\mu,\dirac\nu]$, $\widehat  F_{\mu\nu}(x)$ is the clover discretization of the 
field strength tensor\footnote{Notice that for historical reasons the field strength tensor discretization
$\widehat F_{\mu\nu}$ adopted in Eq.~(\ref{eq:DiracSW}) is not traceless at variance of the one used to define
the energy-momentum tensor, see Eq.~(\ref{eq:FmunuL}).}
\begin{equation}
\label{eq:CloverFmunu}
\widehat  F_{\mu\nu}(x) = {i\over 8a^2}\big\{Q_{\mu\nu}(x)-Q_{\nu\mu}(x)\big\}\; ,
\end{equation}
and  
\begin{equation}
\begin{split}
	Q_{\mu\nu}(x) &= U_\mu(x)U_\nu(x+a \hat{\mu})U^\dag_\mu(x+a \hat{\nu})U^\dag_\nu(x)\\
	&+ U_\nu(x)U_\mu^\dag(x-a \hat{\mu}+a \hat{\nu})U^\dag_\nu(x-a \hat{\mu})U_\mu(x-a \hat{\mu})\\
	&+ U_\mu^\dag(x-a \hat{\mu})U_\nu^\dag(x-a \hat{\mu}-a \hat{\nu})U_\mu(x-a \hat{\mu}-a \hat{\nu})U_\nu(x-a \hat{\nu})\\
	&+ U_\nu^\dag(x-a \hat{\nu})U_\mu(x-a \hat{\nu})U_\nu(x+a \hat{\mu}-a \hat{\nu})U_\mu^\dag(x)\; .
\end{split}
\end{equation}
The coefficient $c_{\rm sw}$ is tuned in order to remove O($a$) lattice artifacts generated by the
action in on-shell correlation functions~\cite{Sheikholeslami:1985ij,Luscher:1996sc}. The gauge-invariant path integral is
\be
Z = \int D U\, D \psibar \, D \psi \; e^{-S}\; . 
\ee
It is also useful to define
\begin{equation}\label{eq:DbfL}
  \nabdbar\mu = \nab\mu - \nabbar\mu\; ,
  \qquad
  \nabdbarstar\mu = \nabstar\mu - \nabbarstar\mu\; ,
\end{equation}
with $\nab\mu,\nabstar\mu$ being the lattice covariant derivatives in Eq.~(\ref{eq:fwd-nablas}), and
\ba
  \nonumber
  a  \psibar(x)\nabbar\mu &= &\psibar(x+a \hat{\mu})U^\dag_\mu(x) - \psibar(x)\; ,\\[0.25cm]
  a \psibar(x)\nabbarstar\mu & = &\psibar(x)-\psibar(x-a \hat{\mu})U_\mu(x-a \hat{\mu})\; . 
\ea

\section{Automatic O($a$)-improvement with massless quarks\label{app:autOa}}
In the absence of spontaneous chiral symmetry breaking, chirally symmetric correlators of Wilson fermions
in the presence of an even number of massless quarks are proven to be
automatically O($a$)-improved~\cite{Frezzotti:2003ni,Sint:2010eh}.
This occurs, for instance, in a finite volume without boundaries~\cite{Sint:2010eh} or in the
thermal theory at high temperature~\cite{DallaBrida:2017sxr}. To extend this result to a
generic number of flavours $N_f>1$ but still small enough to have asymptotic freedom,
we consider the discrete axial symmetry $S_{5}$ defined as
\be\label{eq:g5anom}
\displaystyle \psi \rightarrow \psi' = e^{i {\pi\over\;\, N_f} \gamma_5}\psi\; , \qquad
\psibar \rightarrow \psibar\, ' = \psibar\, e^{i {\pi\over\;\, N_f} \gamma_5}\; .
\ee
This non-anomalous element of the $U(1)_A$ group indeed allows for a simple generalization
of the line of argumentation
given in Refs.~\cite{Frezzotti:2003ni,Sint:2010eh}.
In the massless limit, the action of the Symanzik's effective continuum theory reads 
\be
S_{\rm eff} = S_0 + a S_1 + {\rm O}(a^2)\; ,
\ee
where $S_0$ is defined as in Eq.~(\ref{eq:SQCD}) but with the bare coupling replaced by
the renormalized one, and 
\be
S_1 = c_1 \int d^4 x\,  \psibar(x) \sigma_{\mu\nu} F_{\mu\nu}(x) \psi(x)\; . 
\ee
The leading discretization effects in the connected correlation function of a multi-local
renormalized field $O$ are then given by the corresponding continuum correlation functions with the insertion
either of $S_1$ or of the O($a$)-counterterm $\delta O$ for the field $O$,
\be\label{eq:symobs}
\langle O \rangle^{\rm lat}_{\rm con} = \langle O \rangle_{\rm con} -
a \langle S_1 O \rangle_{\rm con} + a \langle \delta O \rangle_{\rm con} + {\rm O}(a^2) \; . 
\ee
If we restrict ourselves to fields $O_{\rm inv}$  invariant under parity and the chiral symmetry $S_5$,
such as (the fermionic component of) $T_{\mu\nu}$ in Eq.~(\ref{eq:Tmunu}) in the massless limit, the invariance of the measure
and of the action $S_0$ implies
\be
\Big[1- \cos{\Big({2\pi\over N_f}\Big)}\Big]\, \langle S_1 O_{\rm inv} \rangle_{\rm con} = 0 
\ee
when spontaneous chiral symmetry breaking is absent. Moreover $\langle \delta O \rangle_{\rm con}=0$ as well
because fields with the same quantum numbers as $O_{\rm inv}$ but of one dimension higher are
not invariant under the symmetry $S_5$ since they must have and extra derivative and therefore an
extra $\gamma$-matrix with respect to $O_{\rm inv}$, a fact that was essential to prove the
automatic O($a$)-improvement in the twisted mass QCD regularization~\cite{Frezzotti:2003ni}. The
Eq.~(\ref{eq:symobs}) then reads
\be
\langle O_{\rm inv} \rangle^{\rm lat}_{\rm con} = \langle O_{\rm inv} \rangle_{\rm con} + {\rm O}(a^2) \; . 
\ee
The very same conclusion can be reached by using the $R_5$ discrete symmetry in Ref.~\cite{Frezzotti:2003ni}
for $N_f=2$, and an element of the axial subgroup $Z_{N_f}$ of the non-Abelian chiral symmetry group for larger
$N_f$.

\section{Propagators and vertices for perturbation theory\label{eq:appD}}
On the lattice, perturbation theory is normally set-up
in terms of algebra-valued fields $A_\mu(x)$ defined as 
\begin{equation}
 \label{eq:PT}
 U_\mu(x) = e^{-i a g_0A_\mu(x)}=1-i a g_0\,A_\mu(x)+\ldots\; , \qquad
A_\mu(x) = A_\mu^a(x)\, T_a\; , 
\end{equation}
where we opt for a minus sign in the exponential so to recover, in the
naive continuum limit, the widely used conventions in appendix~\ref{App:cont}.
By inserting (\ref{eq:PT}) in the expressions in appendix \ref{App:lat}
and by expanding to the appropriate order, the analogous continuum formulas
given in appendix \ref{App:cont} are recovered after the usual field rescaling
$A_\mu(x) \rightarrow A_\mu(x)/g_0$. The free theory in recovered in the limit
$g_0\rightarrow 0$.

In the presence of the boundary conditions (\ref{eq:shift_lat}), the Fourier
transform can be written as \footnote{To avoid burdening the  notation, we use
the same symbol for the field and for its Fourier transform since any ambiguity
is resolved from the context.}
\be
A_\mu(x) = \int_{p_{_{\vec\xi}}} A_\mu(p)\, e^{ip (x + \frac{a}{2} \hat \mu )}\; , 
\ee
where, for a generic function $f(p)$,
the finite-volume integration is defined to be
\be\label{eq:intB}
\int_{p_{_{\vec\xi}}} f(p) = {1\over{L_0L_1L_2L_3}} \sum_{n} f(p)\; , 
\ee
with
\begin{equation}
\label{eq:MOMbos1}
p_0={2\pi n_0\over L_0}-\sum_{k=1}^3p_k\xi_k\; , 
\qquad
p_k={2\pi n_k\over L_k}\; ,
\end{equation}
and $n_\mu=0,\ldots,L_\mu/a-1$. For lattice fermions satisfying the boundary
conditions in Eqs.~(\ref{eq:psibcs}) and (\ref{eq:psibcs_finiteV}), the Fourier
transform can be written as 
\be\label{eq:FTf}
\psi(x) =  \int_{p_{_{\vec\xi,\theta}}} \psi(p)\, e^{ip x}\; ,\qquad \psibar(x) =  \int_{p_{_{\vec\xi,\theta}}} \psibar(p)\, e^{-ip x}\; ,
\ee
where the integration $\int_{p_{_{\vec\xi,\theta}}}$ is defined as in Eq.~(\ref{eq:intB}) but for
the set of lattice momenta
\be\label{eq:MOMferm1}
p_0 =
{2\pi n_0\over L_0}+{\theta_0  \over L_0} + {\pi\over L_0}-\sum_{k=1}^3 p_k \xi_k\; ,
  \qquad
p_k =  {2\pi n_k\over L_k}+{\theta_k \over L_k}\; . 
\ee
The term $\pi/L_0$ in $p_0$ is due to the anti-periodicity of the boundary conditions along
the temporal direction.

When $L_k\rightarrow\infty$ for all three spatial directions $k$, the integration
in Eq.~(\ref{eq:intB}) becomes
\be\label{eq:infV}
\int_{p_{_{\vec\xi}}} f(p) \rightarrow {1 \over L_0} \sum_{n_0} \int_{BZ}
\displaystyle {d^3 \vec p \over (2\pi)^3}  f(p) \; ,
\ee
where $BZ$ stands for the Brillouin zone. The analogous holds for the fermion integrals 
(\ref{eq:FTf}) which become independent on $\theta_k$. If also $L_0\rightarrow\infty$, the sum
over $n_0$ in Eq.~(\ref{eq:infV}) is replaced by the integral as  well, which becomes independent
on the shift $\vec\xi$ and the twist $\theta_0$.

As expected by general quantum field theory arguments, the expressions of the propagators
and of the vertices of the theory with shifted and twisted boundary conditions are equal to
those valid for periodic boundary conditions provided the definition of the momenta are
replaced by those in Eqs.~(\ref{eq:MOMbos1}) and (\ref{eq:MOMferm1}),  i.e. $\vec\xi$ and
$\theta$ enter the values of the allowed momenta only. For consistency and to define our
conventions, however, we report their definitions in the rest of this appendix.

\subsection{Gluon propagator}
By adding the gauge-fixing contribution\footnote{As usual the backward lattice derivative
$\drvstar\mu$ is defined as Eq.~(\ref{eq:fwd-nablas}) but with the link omitted.}
\be
S^{GF} = a^4 \sum_{x} \Tr\left[(\drvstar\mu A_\mu)(x) (\drvstar\mu A_\nu)(x)\right]
\ee
to the gluonic action (\ref{eq:SG}),
the free-gluon propagator in Feynman gauge reads
\begin{equation}
 \label{eq:G}
 \langle A_\mu^a(x)A_\nu^b(y)\rangle_{\bsxi}=
 \int_{p_{_{\vec\xi}}} D^{ab}_{\mu\nu}(p)\, e^{ip(x-y)}\; , 
\end{equation}
where
\be
D^{ab}_{\mu\nu}(p) = {\delta_{ab}\delta_{\mu\nu} \over D_G(p)}\; , \qquad
D_G(p) =  \sum_{\mu=0}^3 \hat p_\mu^2\; ,\qquad \hat p_\mu = { 2 \over a}
\sin\left({a p_\mu \over 2}\right)\; . 
\ee

\subsection{Ghost propagator}
By following the usual Faddeev-Popov (FP) procedure, we add to
the gluonic action the contribution coming from the FP determinant
\be\label{eq:FPL}
S^{FP} = - 2a^4 \sum_{x} \Tr\left[{\overline c}(x) \drvstar\mu \widehat D_\mu c(x)\right]\; , 
\ee
where
\be
\widehat D_\mu c =  [M(A_\mu)]^{-1} \nab\mu c - i g_0\, [A_\mu,c]\; , \qquad
M(A_\mu) = \left(\frac{\displaystyle 1- e^{-i \Phi(A_\mu)}}{\displaystyle i \Phi(A_\mu)}\right)\; , 
\ee
and $\Phi(A_\mu)$ is a matrix in the adjoint representation of the algebra of SU($N_c$) whose
matrix elements are $[\Phi(A_\mu)]^{ab} = i a g_0 f^{abc} A_\mu^c$. The ghost propagator then reads
\be
  \label{eq:Ghost}
  \langle c^a(x) {\overline c}^b(y) \rangle_{\bsxi} = \int_{p_{_{\vec\xi}}} \Delta^{ab}_G (p)\, e^{ip(x-y)}\; ,
  \qquad \Delta_G (p) = {\delta_{ab} \over D_G(p)}\; .
\ee
Notice that the ghost and anti-ghost fields satisfy the same boundary conditions
of the gauge field.
\subsection{Fermion propagator}
The free-fermion propagator for a single flavour is given by
\be
\langle \psi(x)\psibar(y)\rangle_{\bsxi,\theta} = 
\int_{p_{_{\vec\xi,\theta}}} S(p)\, e^{ip(x-y)}\; , 
\ee
where
\be
S(p) =  {-i \gamma_\mu \, \bar p_\mu  + {m}_0(p) \over D_F(p)}\; , 
\ee
and
\be
D_F(p) = \sum_{\mu=0}^3\bar p_\mu^2 + m_0^2(p)\,, \qquad  m_0(p) =  m_0 + {a \over 2}
\sum_{\mu=0}^3 \hat p^2_\mu\; ,\qquad \bar p_\mu= {1 \over a}\sin( a p_\mu) \; .  
\ee

\subsection{Gluonic interaction}
The perturbative expansion of the Wilson action (\ref{eq:SG}) can be written as
\begin{equation}
 S^G= S^{G,0} + g_0 S^{G,1} + g_0^2 S^{G,2} + {\rm O}(g_0^3)\; , 
\end{equation}
where $S^{G,0}$ is the tree-level gluonic action, while the contributions from the three- and
the four-gluon vertices are
\ba
& & S^{G,1} =  {i\over 6} f_{abc} \int_{k_{\vec\xi};p_{\vec\xi};q_{\vec\xi}} \!\!\!\!\!\!\!\!\!\!\!
\bar\delta(k+p+q) A_\mu^a(k) A_\nu^b(p) A_\lambda^c(q)\, \cdot\\[0.25cm]   
& &\hspace{-1.0cm}\cdot \Big[ \delta_{\lambda\nu}   (\widehat{q-p})_\mu\, \cn\!\Big({ k\over 2}\Big) 
+\delta_{\mu\lambda}   (\widehat{k-q})_\nu\, \cm\!\Big({ p\over 2}\Big)
+\delta_{\mu\nu} (\widehat{p-k})_\lambda\, \cm\!\Big({ q\over 2}\Big)
\Big]\; , \nonumber  
\ea
\begin{equation}
S^{G,2}\!\! =\! {-1\over 24}\!\!\int_{k_{\vec\xi};q_{\vec\xi};r_{\vec\xi};s_{\vec\xi}}\!\!\!\!\!\!\!\!\!\!\!\!\!\!\!\!\!\!\!\!\!
\bar\delta(k\!+\!q\!+\!r\!+\!s)
A_\mu^a(k) A_\nu^b(q) A_\lambda^c(r) A_\rho^d(s)
\big[ {\cal X}^{abcd}_{\mu\nu\lambda\rho}(k,\!q,\!r,\!s)\!+\!{\cal Y}^{abcd}_{\mu\nu\lambda\rho}(k,\!q,\!r,\!s) \big]\; ,  
\end{equation}
with
\ba
&{\cal X}^{abcd}_{\mu\nu\lambda\rho}(k,q,r,s) = -f_{abe} f_{cde} \bigg\{
\delta_{\mu\lambda}\delta_{\nu\rho} \left[ \cm\!\left({q-s\over2}\right) \cn\!\left({k-r\over2}\right) 
  - {{a^4} \over 12} \hat k_\nu \hat q_\mu \hat r_\nu \hat s_\mu \right] \nonumber\\[0.25cm]
&-\delta_{\mu\rho}\delta_{\nu\lambda} \left[ \cm\left({q- r\over2}\right)\! \cn\left({k-s\over2}\right)
-{{a^4} \over 12} \hat k_\nu \hat q_\mu \hat r_\mu \hat s_\nu\right]\!
+\! {a^2\over6}\! \Big[\delta_{\nu\lambda} \delta_{\nu\rho} (\widehat{s-r})_\mu \hat k_\nu \cm({ q \over 2})
 \\[0.25cm]
&- \delta_{\mu\lambda} \delta_{\mu\rho}   (\widehat{s-r})_\nu \hat q_\mu \cn({ k \over 2})
 + \delta_{\mu\nu}    \delta_{\mu\rho}   (\widehat{q-k})_\lambda \hat r_\rho \cl({ s \over 2})
 - \delta_{\mu\nu}    \delta_{\mu\lambda} (\widehat{q-k})_\rho \hat s_\lambda \crh({ r \over 2})\Big]
\nonumber\\[0.25cm] 
&+{a^2\over 12} \delta_{\mu\nu} \delta_{\mu\lambda} \delta_{\mu\rho}\sum_\sigma  (\widehat{q-k})_\sigma (\widehat{s-r})_\sigma
\bigg\} 
\!+\!(b\!\leftrightarrow\! c,\nu\! \leftrightarrow\! \lambda, q\! \leftrightarrow\! r) 
\!+\!(b\!\leftrightarrow\! d,\nu\! \leftrightarrow\! \rho, q\! \leftrightarrow\! s)\, , 
\nonumber
\ea
\ba
&&\hspace{-0.5cm}{\cal Y}^{abcd}_{\mu\nu\lambda\rho}(k,{ q},{ r},{ s})\! =\! {a^4 \over 12} \bigg\{
{2\over N} \big( \delta_{ab}\delta_{cd}\! +\! \delta_{ac}\delta_{bd}\!+\! \delta_{ad}\delta_{bc} \big)\! +\!
\big( d_{abe}d_{cde}\! +\! d_{ace}d_{bde}\!+\! d_{ade}d_{bce} \big)\! \bigg\} \cdot\nonumber \\
& &\hspace{-0.325cm}\bigg\{ 
   \delta_{\mu\nu} \delta_{\mu\lambda} \delta_{\mu\rho} \!\sum_\sigma \hat k_\sigma \hat q_\sigma \hat r_\sigma \hat s_\sigma 
 - \delta_{\mu\nu} \delta_{\mu\lambda} \hat k_\rho \hat  q_\rho \hat r_\rho \hat s_\mu   
 - \delta_{\mu\nu} \delta_{\mu\rho}   \hat k_\lambda \hat q_\lambda \hat s_\lambda \hat r_\mu 
 - \delta_{\mu\lambda} \delta_{\mu\rho}\hat k_\nu \hat r_\nu \hat s_\nu  \hat q_\mu\nonumber\\
&& 
 - \delta_{\nu\lambda} \delta_{\nu\rho}\hat q_\mu \hat r_\mu \hat s_\mu \hat k_\nu
 + \delta_{\mu\nu}    \delta_{\lambda\rho}\hat k_\lambda \hat q_\lambda \hat r_\mu \hat s_\mu
 + \delta_{\mu\lambda} \delta_{\nu\rho} \hat k_\nu \hat r_\nu \hat q_\mu \hat s_\mu
 + \delta_{\mu\rho}   \delta_{\nu\lambda} \hat k_\nu \hat s_\nu \hat q_\mu \hat r_\mu \bigg\}\, , 
\ea
$\cm(p) = \cos(a p_\mu)$ and
\be
\bar \delta(p) = (2\pi)^4 \delta^{(4)}(p) = a^4 \sum_x e^{ip x}\; .
\ee
The Jacobian from the Haar integration measure due to the change of variables
(\ref{eq:PT}) can be recast in the form of an extra contribution to the action which reads
\begin{equation}
S^M= g_0^2 S^{M,2} + {\rm O}(g_0^4)\; , \qquad  S^{M,2} = {N \over 24 a^2}\; \delta_{ab} \delta_{\mu\nu}
\int_{k_{\vec\xi};q_{\vec\xi}} \!\!\!\!\!\!\! \bar\delta(k+q)\, A_\mu^a(k) A_\nu^b(p) \; . 
\end{equation}

\subsection{Ghost-gluon interaction} 
The expansion of the FP action (\ref{eq:FPL}) reads
\begin{equation}
 S^{FP}= S^{FP,0} + g_0 S^{FP,1} + g_0^2 S^{FP,2} + {\rm O}(g_0^3)
\end{equation}
where $S^{FP,0}$ is the tree-level term, while 
\begin{equation}
S^{FP,1} = i f_{abc} \int_{k_{\vec\xi};p_{\vec\xi};q_{\vec\xi}} \!\!\!\!\!\!\!\!\!\!\!\!\!\!\! \bar\delta(p+q-k)
\,{\overline c}^a (p) A_\mu^b(q) c^c(k)
\Big[\hat k_\mu \cm \left({p\over 2}\right)\Big]\; , 
\end{equation}
\begin{equation}
S^{FP,2}\!\! =\! - \delta_{\mu\nu}{a^2  \over 24} \big(f_{abe} f_{dce}\!+\!f_{ace} f_{dbe} \big)\!\!\!
\int_{k_{\vec\xi};q_{\vec\xi};r_{\vec\xi};s_{\vec\xi}} \!\!\!\!\!\!\!\!\!\!\!\!\!\!\!\!\!\!\!\!
\bar\delta(k + q + r - s)
{\overline c}^a (r) A_\mu^b(q)  A_\nu^c(k) c^d(s)\, \hat r_\mu \hat s_\mu\; .
\end{equation}

\subsection{Quark-gluon interaction}
By inserting Eq.~(\ref{eq:PT}) into Eq.~(\ref{eq:fermionicAction}) and by expanding
in $g_0$, the fermionic action reads 
\begin{equation}\label{eq:acFexp}
 S^F= S^{F,0} + g_0 S^{F,1} + g_0^2 S^{F,2} +g_0 S^{{SW},1} + g^2_0 S^{{SW},2} + {\rm O}(g_0^3)\; ,
\end{equation}
where $S^{F,0}$ is the tree-level contribution, while 
\begin{align}
 S^{F,1}=-
\sum_{\mu} \int_{q_{\vec\xi};p_{\vec\xi,\theta};r_{\vec\xi,\theta}}\!\!\!\!\!\!\!\!\!\!\!\!\!\!\!\!\!\!\!\!\! 
 \bar\delta(p-q-r) \,\psibar{}(p) A_\mu(q) 
 \Bigg[i \gamma_\mu \,\cm\! \left( {p+r\over2}\right) + {a\over 2} \,(\widehat{p+r})_\mu\Bigg]
 \psi{}(r)\; , 
\end{align}
\begin{align}
 S^{F,2}\!=\! 
 {a\over2}\!
\sum_{\mu}\!\!\int_{q_{\vec\xi};r_{\vec\xi};p_{\vec\xi,\theta};s_{\vec\xi,\theta}}\!\!\!\!\!\!\!\!\!\!\!\!\!\!\!\!\!\!\!\!\!
\!\!\!\!\!\!\!\bar\delta(p\!-\!q\!-\!r\!-\!s)&\,\psibar{}(p) A_\mu(q)A_\mu(r)\!\Bigg[\cm\!\left({p+s\over2}\right)\!\!
 -i {a\over 2} \gamma_\mu (\widehat{p+s})_\mu\!\Bigg]
 \psi{}(s), 
\end{align}
and 
\begin{equation}
 S^{{SW},1}= - i
 {a\over 2}  \,c_{\rm sw}
\sum_{\mu,\nu}  \int_{q_{\vec\xi};p_{\vec\xi,\theta};r_{\vec\xi,\theta}}\!\!\!\!\!\!\!\!\!\!\!\!\!\!\!\!\!\!\! 
\bar\delta(p-q-r)\,\psibar{}(p) A_\mu(q) \sigma_{\mu\nu}\psi{}(r)\,
\Big[\bar q_\nu
\cm\!\left({q\over 2}\right)\Big] \; .
\end{equation}
The $S^{{SW},2}$ has two quark and two gluonic lines, but it does not contribute to the
quantities we are interested in this paper due to its color structure.\\

\section{The free energy density\label{app:freeE}}
In this appendix we report the coefficients of the 1-loop perturbative expansion
of the free-energy density defined in Eqs.~(\ref{eq:pt})--(\ref{eq:floop}). The gluonic and
fermionic tree-level values are 
\begin{equation}\label{eq:f-free}
f^{G(0)} = \int_{p_{_{\vec\xi}}}  \ln{\Big[a^2\, D_G(p)\Big]}\; ,
\qquad
f^{F(0)} = - 2\int_{p_{_{\vec\xi,\theta}}}  \ln{\Big[a^2 D_F(p)\Big]}\; ,
\end{equation}
and the derivative of the fermionic contribution with respect to the
bare mass is
\be\label{eq:DF0}
{\partial f^{F(0)} \over \partial m_0} =  -4 F^{(8)}\; ,
\ee
where $F^{(8)}$ is defined in appendix \ref{integrals}.
The 1-loop contributions result from connected diagrams with no external legs.
The gluonic ones are
\be
\langle S^{G,2}\!\! +\! S^{FP,2}\!\! +\! S^{M,2}\!\!-\! {(S^{G,1})^2\!\! +\! (S^{FP,1})^2\over2} \rangle_{\rm con}\! = \! 
(N_c^2\!-\! 1) \Big[ N_c f^{G(1,N_c)}\! + {1 \over N_c} f^{G(1,{1 \over N_c})}\!\Big]\; ,
\ee
from which we have 
\bea
f^{G(1,N_c)} & = &\!\!\! \Big\{ (B^{(0)})^2  - {1\over2} \sum_\sigma \Big[B^{(0)}-B_\sigma^{(3)} \Big]^2 +\! {1\over 2} a^2 K_1\! +
\! {1 \over 24} a^4 K_2\! -\! {1\over 2 a^2} B^{(0)} \Big\}\; ,\label{eq:fG-1loop}\\[0.25cm]
f^{G(1,{1\over N_c})} & = & {1 \over 2}  \Big\{ \sum_\sigma \Big[B^{(0)}-B_\sigma^{(3)} \Big]^2 +{1\over 8 a^4} \Big\}\; ,
\eea
where
\bea
K_1 & = & \int_{p_{_{\vec\xi}};q_{_{\vec\xi}};k_{_{\vec\xi}}} {\bar\delta(p+q+k) \over D_G(p)D_G(q)D_G(k)}
\sum_\mu \hat p_\mu^2\, \hat q_\mu^2\; ,\\[0.125cm]
K_2 & = & \int_{p_{\vec\xi};q_{\vec\xi};k_{\vec\xi}} {\bar\delta(p+q+k) \over D_G(p)D_G(q)D_G(k)}
\sum_\mu  \hat p_\mu^2\, \hat q_\mu^2\, \hat k_\mu^2\; .
\eea
The fermionic contribution is
\be
\langle S^{F,2} - {(S^{F,1})^2\over2} \rangle_{\rm con} = 
(N_c^2-1) N_f f^{F(1,N_f)}\; , 
\ee
where
\begin{align}\label{eq:fF-1loop}
& f^{F(1,N_f)} =
B^{(0)} \Bigg[ {1\over a^2} - a \Big(a m_0+ 4 \Big) F^{(8)}\Bigg]\!\!
+\!\! \int_{q_{_{\vec\xi}};p_{_{\vec\xi,\theta}};k_{_{\vec\xi,\theta}}}\!\!\!\! {\bar\delta(p-q-k)\over D_G(q)D_F(p)D_F(k)}\times\\
&\Bigg[a m_0(k)\sum_{\sigma} \bar r_\sigma \bar p_\sigma  + a m_0(p)\sum_{\sigma}\bar k_\sigma \bar r_\sigma
-m_0(k) m_0(p)\sum_{\sigma}\cs(r)
+\sum_{\sigma}\bar p_\sigma \bar k_\sigma \big(\cs(r)-3\big)\Bigg]\; ,\nonumber 
\end{align}
and $r=p+k$.

\subsection{O($a$)-improved action \label{sec:fimpr}}
The insertion of the improvement term modifies only $f^{F(1,N_f)}$ so that
\begin{equation}
f^{F(1,N_f)} \longrightarrow f^{F(1,N_f)} + {\cal F}^{F1} + {\cal F}^{F2}
\end{equation}
where
\bea
\langle (- S^{SW,1} S^{F,1})\rangle_{\rm con} & = & (N_c^2-1) N_f\, {\cal F}^{F1}\; ,\\[0.125cm]  
\langle - {1\over2} (S^{SW,1})^2 \rangle_{\rm con} & = & (N_c^2-1) N_f\, {\cal F}^{F2}\; . 
\eea
The expressions of ${\cal F}^{F1}$ and ${\cal F}^{F2}$ in terms of the integrals
defined in appendix \ref{integrals} are
\begin{align}
& {\cal F}^{F1}  =  - {a c_{\rm sw} \over 2} 
  \int_{q_{_{\vec\xi}};p_{_{\vec\xi,\theta}};k_{_{\vec\xi,\theta}}} {\bar\delta(p-q-k)\over D_G(q) D_F(p) D_F(k)}
  \Bigg\{a\! \sum_{\sigma \rho}\!\! \Big[(\bar p_\rho + \bar k_\rho) \bar q_\sigma
    (\bar p_\sigma \bar k_\rho - \bar k_\sigma \bar p_\rho)\Big] \nonumber\\[0.25cm]
& - \sum_\sigma \Big\{\bar q_\sigma \big[ m_0(k) \bar p_\sigma - m_0(p) \bar k_\sigma \big]
\sum_{\rho\neq\sigma} \big[ \crh (p) + \crh (k) \big]\Big\}\Bigg\}\; , 
\end{align}
and
\begin{align}
& {\cal F}^{F2} =  {a^2 c_{\rm sw}^2\over 8}
\int_{q_{_{\vec\xi}};p_{_{\vec\xi,\theta}};k_{_{\vec\xi,\theta}}}\!\!\!\!\!\!
    {\bar\delta(p-q-k)\over D_G(q) D_F(p) D_F(k)}
    \Bigg\{m_0(p) m_0(k) \sum_\sigma \Big\{{\bar q}^2_\sigma \Big[3 + \sum_{\rho\neq\sigma} \crh (q) \Big]\!\Big\} 
\nonumber\\[0.125cm]
& 
+ 2 \sum_{\sigma\rho} \bar k_\sigma  \bar q_\sigma \bar p_\rho \bar q_\rho 
\Big(2\! -\! \rmc_\sigma(q)\!  +\! \sum_{\lambda\neq\rho}\rmc_\lambda (q) \Big)\! -\!
\sum_{\sigma\rho} \bar k_\sigma  \bar p_\sigma {\bar q}_\rho^2 \Big(1\! -\! 2 \rmc_\sigma(q) +
\!\sum_{\lambda\neq\rho}\rmc_\lambda (q) \Big) 
\Bigg\}\; .
\end{align}
\vspace{0.5cm}

As an explicit check of the whole computation, we compared the derivative of the
free-energy density with respect to $\theta_\mu$ to the expectation value of the
$\mu$-component of the conserved vector current (\ref{eq:VectorCurrent}), e.g.
for $\mu=0$ this corresponds to verify the lattice analog of Eq.~(\ref{eq:dthet1}).
The required 1-loop computation of the expectation value of the vector current is
reported in the following appendix.

\section{Expectation value of the vector current \label{app:freeV}}
By expanding the conserved current in Eq.~(\ref{eq:VectorCurrent}) to order $g_0^2$
\begin{equation}
V^c_\mu = V_\mu^{(0)} + g_0 V_\mu^{(1)} +  g_0^2 V_\mu^{(2)} + {\cal O} (g_0^3)\; , 
\end{equation}
its expectation value at one loop can be written as 
\be\label{eq:ptV}
\langle V^c_\mu \rangle = {\cal V}_\mu^{(0)} +  g_0^2\, {\cal V}_\mu^{(1)}\; . 
\ee
The tree-level value is given by (no summation over $\mu$)
\begin{equation}
{\cal V}_\mu^{(0)} = \langle V_\mu^{(0)} \rangle= 4 i\, N_c N_f \left\{F_{\mu\mu}^{(4)} + a F_\mu^{(5)} \right\}\; , 
\end{equation}
and its derivative with respect to the bare mass is
\be
{\partial {\cal V}_\mu^{(0)}  \over \partial m_0} = 4i \, N_c N_f
\Bigg\{
{a^2 F_{\mu}^{(5)}+\sum_\sigma  aF_{\mu\sigma}^{(4)}\over (a m_0+4)}
-2 \Big( F_{\mu\mu}^{(6)} + a F_\mu^{(7)} \Big) \Bigg\}\; . 
\ee
The 1-loop contribution is
\be
{\cal V}_\mu^{(1)} = i (N_c^2-1) N_f\Big\{{\cal V}_\mu^1 + {\cal V}_\mu^2 + {\cal V}_\mu^3 \Big\}\; , 
\ee
where
\bea
\langle V_\mu^{(2)}\rangle_{\rm con} & = & i (N_c^2-1) N_f {\cal V}_\mu^1  \; ,\\
\langle V_\mu^{(1)} (-S^{F,1}) \rangle_{\rm con} & = & i (N_c^2-1) N_f {\cal V}_\mu^2 \;,\\
\langle V_\mu^{(0)} \Big[ {1\over2} (S^{F,1})^2 - S^{F,2}\Big] \rangle_{\rm con} & = &
i (N_c^2-1) N_f {\cal V}_\mu^3\; .
\eea
The expressions of ${\cal V}_\mu^1$, ${\cal V}_\mu^2$, and ${\cal V}_\mu^3$ in terms of the integrals
defined in appendix \ref{integrals} are
\begin{equation}
{\cal V}_\mu^1 = 
- a^2 B^{(0)} \Big\{F_{\mu\mu}^{(4)}+a F_{\mu}^{(5)}\Big\}\; ,  
\end{equation}
\begin{align}
&{\cal V}_\mu^2 =
- 2 a \int_{q_{_{\vec\xi}};p_{_{\vec\xi,\theta}};k_{_{\vec\xi,\theta}}}
{\bar\delta(p-q-k)\over D_G(q)D_F(p)D_F(k)}\times\\[0.25cm] 
&\Big\{a \bar r_\mu
\big[m_0(k) m_0(p) - \bar k_\mu \bar p_\mu\big] + \cm(r)
  \big[m_0(p) \bar k_\mu + m_0(k) \bar p_\mu \big]\Big\}\; ,\nonumber
\end{align}
and 
\begin{align}
& {\cal V}_\mu^3 =
B^{(0)} \Bigg\{a^2 F_{\mu\mu}^{(4)}+ 2 a^3 F_{\mu}^{(5)} + a^2 \sum_{\sigma} F_{\mu\sigma}^{(4)}
-2 (a m_0+4) \Big[a F_{\mu\mu}^{(6)}+ a^2 F_{\mu}^{(7)} \Big]  \Bigg\}\nonumber  \\
&-2  \int_{q_{_{\vec\xi}};p_{_{\vec\xi,\theta}};k_{_{\vec\xi,\theta}}}\!\!\! {\bar\delta(p-q-k)\over D_G(q)D^2_F(k)D_F(p)}
\Bigg\{ 2 \bar k_\mu  \big[\cm(k)+ a m_0(k)\big] \Bigg\{m_0(p)m_0(k)\sum_{\sigma}\cs(r) \nonumber \\
&  - a \sum_{\sigma} \Big\{\bar r_\sigma \Big[ m_0(k) \bar p_\sigma
  + m_0(p) \bar k_\sigma \Big]\Big\} +
\sum_{\sigma} \Big\{\bar p_\sigma \bar k_\sigma \Big[3-\cs(r)\Big]\Big\} \Bigg\}\\
 & + D_F(k) \Bigg\{a m_0(p) \bar r_\mu \cm(k)+ \bar p_\mu \cm(k)\big[\cm(r)-3\big] -
a \bar k_\mu\Bigg[m_0(p)\sum_{\sigma}\cs(r) -
a \sum_{\sigma} \bar r_\sigma \bar p_\sigma\Bigg]\Bigg\}\Bigg\}\nonumber
\end{align}
where we have defined $ r =  p +  k$.

\subsection{O($a$)-improved action \label{sec:V0impr}}
The Sheikholeslami-Wohlert terms leads to 3 additional terms to the 1-loop coefficient
\begin{equation}
{\cal V}_\mu^{(1)} \longrightarrow {\cal V}_\mu^{(1)} +
i (N_c^2-1) N_f\Big\{{\cal V}_\mu^4 + {\cal V}_\mu^5 + {\cal V}_\mu^6 \Big\}
\end{equation}
where
\bea
\langle V_\mu^{(0)} S^{SW,1} S^{F,1}\rangle_{\rm con} & = & i (N_c^2-1) N_f {\cal V}_\mu^4\; , \\
\langle V_\mu^{(0)} {(S^{SW,1})^2\over 2} \rangle_{\rm con}   & = & i (N_c^2-1) N_f {\cal V}_\mu^5\; , \\
\langle V_\mu^{(1)} (-S^{SW,1}) \rangle_{\rm con} & = & i (N_c^2-1) N_f {\cal V}_\mu^6\; . 
\eea
The expressions of ${\cal V}_\mu^4$, ${\cal V}_\mu^5$, and ${\cal V}_\mu^6$ in terms of the integrals defined
in appendix \ref{integrals} are
\begin{align}
  & {\cal V}_\mu^4 = a c_{\rm sw} 
  \int_{q_{_{\vec\xi}};p_{_{\vec\xi,\theta}};k_{_{\vec\xi,\theta}}} {\bar\delta(p-q-k)\over D_G(q) D^2_F(k)D_F(p)}
  \Bigg\{ 2 \bar k_\mu \Big[ \cm (k) + a m_0(k) \Big] \Bigg\{ \nonumber \\
  & a \sum_{\sigma\rho}\Big\{ \bar q_\sigma (\bar k_\sigma \bar p_\rho -\bar p_\sigma \bar k_\rho) \big(\bar p_\rho + \bar k_\rho \big)\Big\}
+ \sum_\sigma \Big\{ \bar q_\sigma \big[m_0(k) \bar p_\sigma - m_0(p) \bar k_\sigma \big] \sum_{\rho\neq\sigma} \big[ \crh (p) + \crh (k) \big]\Big\}\Bigg\}
\nonumber\\
& + D_F(k) 
\Bigg\{
\cm(k) \Bigg\{a \big(\bar p_\mu + \bar k_\mu \big) \sum_\sigma \bar q_\sigma \bar p_\sigma
+ \bar q_\mu  \Big[
m_0(p) \sum_{\sigma\neq\mu} \Big( \cs (p) + \cs (k) \Big)\\
&  
-a \sum_\sigma \bar p_\sigma \big(\bar p_\sigma + \bar k_\sigma \big)
\Big]\Bigg\}
   - a \bar k_\mu \sum_\sigma \Big[\bar q_\sigma \bar p_\sigma \sum_{\rho\neq\sigma} \big( \crh (p) + \crh (k) \big)\Big] \Bigg\}\; ,  \nonumber
\end{align}
\begin{align}
&{\cal V}_\mu^5 =
{a^2 c_{\rm sw}^2 \over4} \int_{q_{_{\vec\xi}};p_{_{\vec\xi,\theta}};k_{_{\vec\xi,\theta}}}
{\bar\delta(p-q-k) \over D_G(q) D^2_F(k)D_F(p)} \Bigg\{
2 \bar k_\mu \Big[ \cm (k) + a m_0(k) \Big]\times\\
&\Bigg\{
2 \sum_\sigma \bar q^2_\sigma
\sum_\rho \bar p_\rho \bar k_\rho \big( 1+\crh (q) \big)
+2 \sum_\sigma \bar k_\sigma \bar q_\sigma 
\sum_\rho \bar q_\rho \bar p_\rho \Big( 2 -\cs (q) +\sum_{\lambda\neq\rho} \rmc_\lambda (q) \Big)\nonumber \\
&
-\Big[ \sum_\sigma \bar p_\sigma \bar k_\sigma -m_0(k) m_0(p) \Big]
\Big[ \sum_\rho  \bar q^2_\rho \Big( 3+\sum_{\lambda\neq\rho} \rmc_\lambda (q) \Big)\Big] 
\Bigg\}\nonumber\\
&+ D_F(k) \Bigg\{ \cm(k)\Big\{\bar p_\mu \sum_\sigma \Big[ \bar q^2_\sigma
  \Big( 1- 2 \cm (q) +\sum_{\rho\neq\sigma} \crh (q) \Big)\Big]
-2 \bar q_\mu 
\sum_\sigma \Big[ \bar p_\sigma \bar q_\sigma \times
  \nonumber \\ &
\Big( 2-\cm(q)+\sum_{\rho\neq\sigma} \crh (q) \Big)\Big]\Big\}
- a m_0(p) \bar k_\mu \sum_\sigma \Big\{ \bar q^2_\sigma  \Big( 3+\sum_{\rho\neq\sigma} \crh (q) \Big)  \Big\}\Bigg\} 
\Bigg\} \nonumber \; , 
\end{align}
and 
\begin{align}
& {\cal V}_\mu^6=
{a^2 c_{\rm sw} \over 2}  \int_{q_{_{\vec\xi}};p_{_{\vec\xi,\theta}};k_{_{\vec\xi,\theta}}} 
{\bar\delta(p-q-k)\over D_G(q) D_F(p) D_F(k)} \Bigg\{
a \Big[\bar p_\mu + \bar k_\mu \Big]\times \\
& \Big[m_0(k) \sum_{\sigma\neq\mu} \bar p_\sigma \bar q_\sigma  - m_0(p)
  \sum_{\sigma\neq\mu} \bar k_\sigma \bar q_\sigma\Big]
-\Big[ \cm (p) + \cm(k) \Big] 
\sum_\sigma \Big[\bar q_\sigma (\bar p_\mu \bar k_\sigma - \bar k_\mu \bar p_\sigma) \Big] \Bigg\}\; .\nonumber 
\end{align}

\section{Expectation values of sextet components of $T_{\mu\nu}$ \label{app:freeT6}}
In this appendix we report the coefficients of the 1-loop perturbative expansion
of the expectation values of the sextet components of the energy-momentum tensor
defined in Eqs.~(\ref{eq:Tsextet})--(\ref{eq:T1lsextet}). By expanding  the field
strength $F_{\mu\nu}$ to order $g_0^2$
\begin{equation}
F_{\mu\nu}^a (x) = F_{\mu\nu}^{a(0)} (x) + g_0 F_{\mu\nu}^{a(1)} (x) +  g_0^2
F_{\mu\nu}^{a(2)} (x) + {\cal O} (g_0^3)\; , 
\end{equation}
the tree-level value of the gluonic part is given by
\be
{\cal T}^{G\{6\}(0)}_{\mu\nu} = {1 \over N_c^2-1} (1-\delta_{\mu\nu})
\sum_{\alpha\neq \mu,\nu} \langle F_{\mu\alpha}^{a(0)} F_{\nu\alpha}^{a(0)} \rangle
= (1-\delta_{\mu\nu}) \sum_{\alpha\neq \mu,\nu} B^{(2)}_{\mu\nu\alpha}\; .   
\ee
The 1-loop contributions to the gluonic component are 
\bea
{\cal T}^{G\{6\}(1,N_c)}_{\mu\nu} & = &(1-\delta_{\mu\nu})\Big\{ {\cal T}^{G1}_{\mu\nu} + {\cal T}^{G2}_{\mu\nu} + {\cal T}^{G3}_{\mu\nu}\Big\} \; ,  \\
{\cal T}^{G\{6\}(1,{1 \over N_c})}_{\mu\nu} & = & (1-\delta_{\mu\nu})\Big\{
{\cal T}^{G4}_{\mu\nu} + {\cal T}^{G5}_{\mu\nu}\Big\} \; , \\
{\cal T}^{G\{6\}(1,N_f)}_{\mu\nu} & = & (1-\delta_{\mu\nu})\Big\{
{\cal T}^{G6}_{\mu\nu} + {\cal T}^{G7}_{\mu\nu}
\Big\}\; , 
\eea
where 
\begin{align}
& \sum_{\alpha\neq \mu,\nu} \langle \big[ 
F_{\mu\alpha}^{a(0)} F_{\nu\alpha}^{a(2)} 
+ F_{\mu\alpha}^{a(2)} F_{\nu\alpha}^{a(0)} 
+ F_{\mu\alpha}^{a(1)} F_{\nu\alpha}^{a(1)}
\big]\rangle_{\rm con} = (N_c^2\!-\!1)\Big( N_c\, {\cal T}^{G1}_{\mu\nu}\!  +\! {1\over N_c} {\cal T}^{G4}_{\mu\nu}\Big)\, , \nonumber\\
& \sum_{\alpha\neq \mu,\nu} \langle \big[
  F_{\mu\alpha}^{a(0)} F_{\nu\alpha}^{a(1)}
+ F_{\mu\alpha}^{a(1)} F_{\nu\alpha}^{a(0)} 
\big](-S^{G,1}) \rangle_{\rm con} = (N_c^2-1) N_c\, {\cal T}^{G2}_{\mu\nu}   \; ,\nonumber \\
& \sum_{\alpha\neq \mu,\nu}\!\!\!\! \langle \big[
  F_{\mu\alpha}^{a(0)} F_{\nu\alpha}^{a(0)}
  \big]\! ({{(S^{G,1})^2\!\!\! +\!\! (S^{FP,1})^2}\over2}\!-\! S^{G,2}\!\!- \! S^{FP,2}\!\!- \!S^{M,2})\! \rangle_{\rm con}\!\! =\!
(N_c^2\!\!-\! 1)\!\Big( N_c {\cal T}^{G3}_{\mu\nu}\!\! +\!\! {1 \over N_c} {\cal T}^{G5}_{\mu\nu}\Big)\,,\nonumber\\
& \sum_{\alpha\neq \mu,\nu} \langle \big[
  F_{\mu\alpha}^{a(0)} F_{\nu\alpha}^{a(0)}
  \big](-S^{F,2}) \rangle_{\rm con} = (N_c^2-1) N_f\, {\cal T}^{G6}_{\mu\nu}\; ,\nonumber \\
& \sum_{\alpha\neq \mu,\nu} \langle \big[
F_{\mu\alpha}^{a(0)} F_{\nu\alpha}^{a(0)}
\big]{{(S^{F,1})^2}\over2} \rangle_{\rm con} =  (N_c^2-1) N_f\, {\cal T}^{G7}_{\mu\nu}\; . 
\end{align}
The expressions of ${\cal T}^{G1}_{\mu\nu},\ldots ,{\cal T}^{G7}_{\mu\nu}$ in terms of the integrals
defined in appendix \ref{integrals} are
\begin{align}
&{\cal T}^{G1}_{\mu\nu}\!\! =\! {a^2 \over 8}\!\!\! \sum_{\alpha\neq \mu,\nu}\!\!\! \Big\{
2 [B_\mu^{(3)}\!+\!B_\nu^{(3)}\!-\!B_\alpha^{(3)}\! -\! {22 \over 3} B^{(0)}] B_{\mu\nu\alpha}^{(2)}\! 
-\! a^2 B_{\mu\alpha}^{(4)} B_{\nu\alpha}^{(4)}\!
+\! B_{\mu\nu}^{(4)} [B^{(0)}\! +\! B_{\alpha}^{(3)}] \\
&+\delta_{\mu\nu} \Big[ 2 [B^{(0)}\! +\! B_{\alpha}^{(3)}][B^{(0)}\! +\! B_{\mu}^{(3)}]
-3 B_{\mu\alpha}^{(4)}B_{\mu\alpha}^{(4)}
-4 [B_\mu^{(3)}\!+\!B_{\mu\alpha}^{(6)}\!+\!B_{\alpha\alpha}^{(4)} \!-\!2 B_{\alpha\alpha\mu}^{(2)}] B_{\mu\mu\alpha}^{(2)}
\nonumber \\
&
+ 8 B_{\mu\alpha\alpha}^{(2)}B_{\alpha\mu\mu}^{(2)}
+ [B^{(0)}\! +\! B_{\mu}^{(3)} \!+\! 2 B_{\mu\mu}^{(4)}] B_{\alpha\alpha}^{(4)} 
-4 [{ 1 \over 2} B_\mu^{(3)}\!+\!B_{\mu\alpha}^{(6)}\!+\!B_{\mu\mu}^{(4)}\! +\! {11 \over 3} B^{(0)}] B_{\alpha\alpha\mu}^{(2)}
\Big]\Big\} \nonumber
\end{align}
\begin{align}
&  {\cal T}^{G2}_{\mu\nu} = {1\over2}
\int_{p_{_{\vec\xi}};q_{_{\vec\xi}};k_{_{\vec\xi}}}
{\bar\delta(p+q+k)\over D_G(p)D_G(q)D_G({k})} \sum_{\alpha\neq \mu,\nu}
\Bigg\{a^2 \bar p_\mu \bar p_\nu \bar q_\alpha
\Big[ 2 \bar q_\alpha - \bar k_\alpha - \bar p_\alpha \Big] \\
&-a^2 \bar p_\mu \bar q_\nu \Big(\bar k_\alpha\!\! -\! \bar p_\alpha \Big) \Big(\bar k_\alpha\!\! -\! \bar q_\alpha \Big)\!
-\! {1\over2} (1\!+\!\ca (p))(1\!+\! \ca (k)) 
\Big[\bar p_\mu (\overline{p-q})_\nu\! +\! \bar p_\nu (\overline{p-q})_\mu\Big] \nonumber \\
& +\delta_{\mu\nu}\Big[a^2 \bar p_\mu \bar q_\mu \bar p_\alpha \Big(\bar q_\alpha\!\! -\! \bar p_\alpha \Big) 
-(1\!+\!\cm (p))(1\!+\! \cm (k)) \bar p_\alpha (\overline{p-q})_\alpha \Big] \Bigg\}\, ,\nonumber
\end{align}
\begin{align}
&{\cal T}^{G3}_{\mu\nu}\!\!=\!\!\!\!\!\sum_{\alpha\neq \mu,\nu}\!\!\Bigg[
 {5\over 6}  B^{(0)} (a^2 B_{\mu\nu\alpha}^{(2)}\!\! -\! 3 B_{\mu\nu\alpha}^{(1)})\!
+\!{1\over 4a^2}  B_{\mu\nu\alpha}^{(1)}\!
+\!{1\over2}  B_\alpha^{(3)} (B_{\mu\nu\alpha\alpha}^{(5)}\!\!-\! 6 B_{\mu\nu\alpha}^{(1)}) 
\!+\!\!{1\over 2}\!  \sum_\sigma B_\sigma^{(3)} B_{\mu\nu\alpha\sigma}^{(5)}\Bigg] \nonumber\\
&+{1\over8} \int_{p_{_{\vec\xi}};q_{_{\vec\xi}};k_{_{\vec\xi}}}\!\!\!
{\bar\delta(p+q+k)\over D_G(p)^2 D_G(q)D_G({k})}
\sum_{\alpha\neq \mu,\nu} \Bigg\{ \Big[8- a^2 D_G(p)\Big]
\Bigg[ \bar p_\mu \bar p_\nu \Big[\bar q_\alpha - \bar k_\alpha \Big]^2 \\
&- \bar p_\mu \bar p_\alpha \Big[\bar q_\alpha - \bar k_\alpha \Big]\Big[\bar q_\nu - \bar k_\nu \Big]
-\bar p_\nu \bar p_\alpha \Big[\bar q_\mu - \bar k_\mu \Big]\Big[\bar q_\alpha - \bar k_\alpha \Big]
+ \bar p^2_\alpha \Big[\bar q_\mu - \bar k_\mu \Big]\Big[\bar q_\nu - \bar k_\nu \Big]  \Bigg]\nonumber \\
 &  - 2\, \bar p_\mu \bar p_\nu \Bigg[a^4 
 \bar p^2_\alpha \Big(\hat q_\mu^2 - \hat q_\alpha^2\Big)
\Big(\hat k_\nu^2 - \hat k_\alpha^2\Big)
- \Big( 1+\ca (p)\Big)\Big( 1+ \ca (q) \Big)
\sum_{\sigma} (\widehat{p-k})^2_\sigma \Bigg]  \nonumber\\ 
&+ 2 \delta_{\mu\nu} \bar p_\alpha^2 \Big( 1+\cm (p)\Big)\Big( 1+ \cm (q) \Big)
\sum_{\sigma} (\widehat{p-k})^2_\sigma  \Bigg\} +
\delta_{\mu\nu} \! \sum_{\alpha\neq \mu}\!\Bigg[
{5\over 6}  B^{(0)} (a^2 B_{\alpha\alpha\mu}^{(2)} -3 B_{\alpha\alpha\mu}^{(1)})
\nonumber\\
&+ {1\over 4a^2} B_{\alpha\alpha\mu}^{(1)} + {1\over 2} B_\mu^{(3)} (B_{\alpha\alpha\mu\mu}^{(5)}
- 6 B_{\alpha\alpha\mu}^{(1)}) + {1\over 2} \sum_\sigma B_\sigma^{(3)} B_{\alpha\alpha\mu\sigma}^{(5)}\Bigg]
\, , \nonumber
\end{align}
\bea
{\cal T}^{G4}_{\mu\nu} & =&  a^2 \sum_{\alpha\neq \mu,\nu} \Big\{ 2 B^{(0)} -{1\over2} [B_\mu^{(3)}+B_\nu^{(3)}]
- B_\alpha^{(3)} \Big\} \Big[ B_{\mu\nu\alpha}^{(2)} +\delta_{\mu\nu} B_{\alpha\alpha\mu}^{(2)} \Big] \, ,\\
{\cal T}^{G5}_{\mu\nu} & = &\!\!\!\!\! - {1 \over 2}\!\!\! \sum_{\alpha\neq \mu,\nu}\!\!\! \Bigg\{\!
a^2\! B_{\mu\nu\alpha}^{(2)} (2 B^{(0)}\!\!-\! B_\alpha^{(3)})\! +\! ({1\over a^2}\!-\! 8 B^{(0)}) B_{\mu\nu\alpha}^{(1)}\!\!
+\! 2\! \sum_\sigma B_\sigma^{(3)} B_{\mu\nu\alpha\sigma}^{(5)}\!\! \\
& + & \!\! \delta_{\mu\nu} \Big[
a^2\! B_{\alpha\alpha\mu}^{(2)} \Big(2 B^{(0)}\!\!-\! B_\mu^{(3)}\Big)\! +\! \Big({1\over a^2}\!-\! 8 B^{(0)}\Big) B_{\alpha\alpha\mu}^{(1)}\!\!
+\! 2\! \sum_\sigma B_\sigma^{(3)} B_{\alpha\alpha\mu\sigma}^{(5)} \Big]\Bigg\}\, ,\nonumber
\eea
\bea
{\cal T}^{G6}_{\mu\nu} & = & -2 a \sum_{\alpha\neq \mu,\nu}\Bigg\{
\Big(a F_\alpha^{(2)}-F_\alpha^{(3)} \Big) B_{\mu\nu\alpha}^{(1)} +
\delta_{\mu\nu} \Big(a F_\mu^{(2)}-F_\mu^{(3)} \Big) B_{\alpha\alpha\mu}^{(1)}\Bigg\}\, ,\\
{\cal T}^{G7}_{\mu\nu} & = & -2 \sum_{\alpha\neq \mu,\nu}\Big(
h^{(1)}_{\mu\nu\alpha\alpha} -  h^{(1)}_{\mu\alpha\alpha\nu} - h^{(1)}_{\alpha\nu\mu\alpha} +  h^{(1)}_{\alpha\alpha\mu\nu} \Big)\, ,
\eea
where
\begin{align}
& \!\! h^{(1)}_{\rho\sigma\mu\nu}\!\!\!=\!\!{1\over 4}\!\! \int_{q_{_{\vec\xi}};p_{_{\vec\xi,\theta}};k_{_{\vec\xi,\theta}}}\!\!\!\!\!\!\! 
{\bar\delta(p-q-k)\bar q_\rho \bar q_\sigma \over D_G(q)^2 D_F(p)D_F(k)}\! \Bigg\{
\!\!a^2\Big[\bar p_\mu\! +\! \bar k_\mu \Big]\!\! \Big[\bar p_\nu\! +\! \bar k_\nu \Big]\!\!
\Big[ m_0(p) m_0(k)\!\! -\!\!\!\sum_\sigma\! \bar p_\sigma \bar k_\sigma \Big]\\[0.25cm]
&\hspace{-0.35cm} +\!\!\Big[\cm (p)\! +\! \cm (k) \Big]\! \Big[ \cn (p)\! +\! \cn (k) \Big]\! 
\Big[\bar k_\mu \bar p_\nu\! +\! \bar p_\mu \bar k_\nu \Big]\!\!
+\!\! a \Big[ \cn (p)\! +\! \cn (k) \Big]\! \Big[ \bar p_\mu\! +\! \bar k_\mu \Big]\!
\Big[ m_0(p) \bar k_\nu\! +\! m_0(k) \bar p_\nu \Big] \nonumber\\
&\hspace{-0.35cm} +\!\! a \Big[ \cm (p)\! +\! \cm (k) \Big]\!\! \Big[\bar p_\nu\! +\! \bar k_\nu \Big]\!\! 
\Big[ m_0(p) \bar k_\mu\! +\! m_0(k) \bar p_\mu \Big]\!\!
-\!\delta_{\mu\nu}\! \Big[ \cm (p)\! +\! \cm (k) \Big] ^2\!
\Big[ m_0(p) m_0(k)\! +\!\!\sum_\sigma \bar p_\sigma \bar k_\sigma\Big] 
\!\!\Bigg\}\, .\nonumber
\end{align}
By expanding the fermion part of the energy-momentum tensor in Eq.~(\ref{eq:BareTF}) to order $g_0^2$
\begin{equation}
T^F_{\mu\nu} (x) = T_{\mu\nu}^{F(0)} (x) + g_0 T_{\mu\nu}^{F(1)} (x) +  g_0^2
T_{\mu\nu}^{F(2)} (x) + {\cal O} (g_0^3)\; , 
\end{equation}
the tree-level value of the expectation value of $T^{F, \{6\}}_{\mu\nu}$ is
\begin{equation}
{\cal T}^{F\{6\}(0)}_{\mu\nu} = {1\over N_c N_f}(1-\delta_{\mu\nu})   
\langle T_{\mu\nu}^{F(0)} \rangle = -4 (1-\delta_{\mu\nu}) F_{\mu\nu}^{(0)}\; , 
\end{equation}
and its derivative with respect to the bare mass is
\be\label{eq:DT6}
{\partial {\cal T}^{F\{6\}(0)}_{\mu\nu}  \over \partial m_0} =  8 (1-\delta_{\mu\nu}) F^{(1)}_{\mu\nu}\; . 
\ee
The 1-loop fermion contribution is
\be
{\cal T}^{F\{6\}(1,N_f)}_{\mu\nu} = (1-\delta_{\mu\nu})
\Big\{{\cal T}^{F1}_{\mu\nu} + {\cal T}^{F2}_{\mu\nu} + {\cal T}^{F3}_{\mu\nu}\Big\} \; ,
\ee
where
\bea
\langle T_{\mu\nu}^{F(2)}\rangle_{\rm con} & = &  (N_c^2-1) N_f\,  {\cal T}^{F1}_{\mu\nu}\; , \\[0.25cm]
\langle T_{\mu\nu}^{F(1)} (-S^{F,1}) \rangle_{\rm con} & = & (N_c^2-1) N_f\,  {\cal T}^{F2}_{\mu\nu}\; , \\[0.25cm]
\left\langle T_{\mu\nu}^{F(0)} \Big({(S^{F,1})^2\over2}-S^{F,2}\Big)\right\rangle_{\rm con} & = & (N_c^2-1) N_f\,  {\cal T}^{F3}_{\mu\nu}\; . 
\eea
The expressions of ${\cal T}^{F1}_{\mu\nu}$, ${\cal T}^{F2}_{\mu\nu}$, and ${\cal T}^{F3}_{\mu\nu}$ in terms of integrals
defined in appendix \ref{integrals} are
\begin{equation}
{\cal T}^{F1}_{\mu\nu} = a^2 B^{(0)} F_{\mu\nu}^{(0)}\; , 
\end{equation}
\begin{align}
& {\cal T}^{F2}_{\mu\nu} =
{1\over2} \int_{q_{_{\vec\xi}};p_{_{\vec\xi,\theta}};k_{_{\vec\xi,\theta}}}\!\!\!\! 
{\bar\delta(p-q-k)\over D_G(q)D_F(p)D_F(k)} 
\Bigg\{a \bar r_\nu
\Big[m_0(p) \bar k_\mu + m_0(k) \bar p_\mu \Big]
\nonumber\\[0.125cm]
&  +a \bar r_\mu
 \Big[m_0(p) \bar k_\nu + m_0(k) \bar p_\nu \Big]
+ \Big[2 + \cm\! (r) + \cn\! (r)\Big]
\Big[\bar k_\mu \bar p_\nu + \bar k_\nu \bar p_\mu\Big] \\[0.125cm]
& - 2 \delta_{\mu\nu} \Big[1+\cm\! (r)\Big]
 \Big[m_0(k) m_0(p)+\sum_{\sigma} \bar k_\sigma \bar p_\sigma \Big]\Bigg\}\; ,\nonumber
\end{align}
\begin{align}
&{\cal T}^{F3}_{\mu\nu} = \Bigg\{\Big[ 2a (a m_0+4) B^{(0)} F_{\mu\nu}^{(1)} - a^2 B^{(0)} F_{\mu\nu}^{(0)} \Big]
+\!\!\int_{q_{_{\vec\xi}};p_{_{\vec\xi,\theta}};k_{_{\vec\xi,\theta}}}\!\!\!\! {\bar\delta(p-q-k)\over D_G(q)D^2_F(k)D_F(p)}\times\nonumber\\
& \Bigg\{ 
\!\! 4 \bar k_\mu \bar k_\nu \Bigg[ m_0(p)\! \Big[ m_0(k)\! \sum_{\sigma}c_\sigma(r)
\! -\! a \sum_{\sigma} \bar k_\sigma \bar r_\sigma\Big]\!\!-\! a m_0(k)\!\!\sum_{\sigma} \bar p_\sigma \bar r_\sigma
 + \sum_{\sigma} [3-\cs(r)] \bar p_\sigma \bar k_\sigma \! \Bigg]
\nonumber\\
 &\!\! +\!\!
D_F(k)\! 
\Bigg[a m_0(p)\Big(\bar r_\mu \bar k_\nu\! + \! \bar k_\mu \bar r_\nu\Big) \!
 +\! \bar p_\mu \bar k_\nu \Big(\cm(r)\!-\!3\Big)
\! +\! \bar p_\nu  \bar k_\mu \Big(\cn(r)\!\! -\! 3\Big)\Bigg]
\Bigg\}\Bigg\}\; , 
\end{align}
where $r=p+k$.

\subsection{O($a$)-improved action}
The 1-loop gluonic contribution to the off-diagonal components of the
energy-momentum due to the improvement term is
\be
   {\cal T}^{G\{6\}(1,N_f)}_{\mu\nu} \longrightarrow {\cal T}^{G\{6\}(1,N_f)}_{\mu\nu} +
(1-\delta_{\mu\nu})\Big\{{\cal T}^{G8}_{\mu\nu} + {\cal T}^{G9}_{\mu\nu}\Big\}\; ,   
\ee   
where
\bea
\sum_{\alpha\neq \mu,\nu} \langle \big[F_{\mu\alpha}^{a(0)} F_{\nu\alpha}^{a(0)}
\big](S^{SW,1} S^{F,1})\rangle_{\rm con} & = & (N_c^2-1) N_f\, {\cal T}^{G8}_{\mu\nu}\; , \\
\sum_{\alpha\neq \mu,\nu} \langle \big[F_{\mu\alpha}^{a(0)} F_{\nu\alpha}^{a(0)}\big]{(S^{SW,1})^2\over2}
\rangle_{\rm con} & = & (N_c^2-1) N_f\, {\cal T}^{G9}_{\mu\nu}\; .
\eea
The expressions of ${\cal T}^{G8}_{\mu\nu}$ and ${\cal T}^{G9}_{\mu\nu}$ in terms of the integrals defined in appendix
\ref{integrals} are
\bea
{\cal T}^{G8}_{\mu\nu} & = & - a c_{\rm sw} \sum_{\alpha\neq \mu,\nu}
  \Big( h^{(2)}_{\mu\nu\alpha\alpha} -  h^{(2)}_{\mu\alpha\alpha\nu} - h^{(2)}_{\alpha\nu\mu\alpha} +
  h^{(2)}_{\alpha\alpha\mu\nu} \Big)\; , \\
{\cal T}^{G9}_{\mu\nu} & = & - {a^2 c_{\rm sw}^2 \over 4}  \sum_{\alpha\neq \mu,\nu}
\Big( h^{(3)}_{\mu\nu\alpha\alpha} -  h^{(3)}_{\mu\alpha\alpha\nu} - h^{(3)}_{\alpha\nu\mu\alpha}
+  h^{(3)}_{\alpha\alpha\mu\nu} \Big)\; ,
\eea
where
\begin{align}
&  h^{(2)}_{\rho\sigma\mu\nu}={1\over 4}
\int_{q_{_{\vec\xi}};p_{_{\vec\xi,\theta}};k_{_{\vec\xi,\theta}}} 
{\bar\delta(p-q-k) \bar q_\rho \bar q_\sigma \over D_G(q)^2
  D_F(p)D_F(k)} \Bigg\{\bigg\{
\Big( 1 + \cn (q) \Big) \times \\[0.25cm]
& \Big[
a \Big(\bar p_\mu + \bar k_\mu \Big)
\Big(\bar p_\nu \sum_\sigma \bar q_\sigma \bar k_\sigma   - \bar k_\nu \sum_\sigma \bar q_\sigma \bar p_\sigma\Big)
+\Big( \cm (p) + \cm (k) \Big) \bar q_\mu \Big( m_0(p) \bar k_\nu - m_0(k) \bar p_\nu \Big)
\Big] %
\nonumber \\
&+ (\mu \leftrightarrow \nu)\bigg\}  + 2 \delta_{\mu\nu} \Big( 1 + \cm (q) \Big)  \Big( \cm (p) + \cm (k) \Big)
\Big[ m_0(k) \sum_\sigma \bar q_\sigma \bar p_\sigma - m_0(p) \sum_\sigma \bar q_\sigma \bar k_\sigma \Big]\Bigg\}\, ,  \nonumber
\end{align}
and
\begin{align}
&  h^{(3)}_{\rho\sigma\mu\nu}={1\over 2} \int_{q_{_{\vec\xi}};p_{_{\vec\xi,\theta}};k_{_{\vec\xi,\theta}}}\!\!\! 
{\bar\delta(p-q-k) \bar q_\rho \bar q_\sigma \over D_G(q)^2
D_F(p)D_F(k)} \Big( 1+ \cm (q) \Big) \Big( 1+ \cn (q) \Big)
\Bigg\{ \bar q_\mu \bar q_\nu \sum_\sigma \bar p_\sigma \bar k_\sigma +\nonumber \\[0.25cm] 
& \Big( \bar k_\mu \bar p_\nu\! +\! \bar k_\nu \bar p_\mu \Big)\!\! \sum_\sigma \bar q^2_\sigma\! 
-\! m_0(k) m_0(p) \bar q_\mu \bar q_\nu 
\! - \! \Big(\bar k_\mu \bar q_\nu\! +\! \bar k_\nu \bar q_\mu \Big)\!\! \sum_\sigma \bar q_\sigma \bar p_\sigma\!
-\!\!\Big(\bar p_\mu \bar q_\nu\! +\! \bar p_\nu \bar q_\mu \Big)\! \sum_\sigma \bar q_\sigma \bar k_\sigma 
\nonumber\\
& +\delta_{\mu\nu} \Big[
  \Big( m_0(k) m_0(p) - \sum_\rho \bar k_\rho \bar p_\rho \Big) \sum_\sigma \bar q^2_\sigma 
  + 2 \Big( \sum_\sigma \bar q_\sigma \bar k_\sigma \Big)
  \Big( \sum_\rho \bar q_\rho \bar p_\rho  \Big)\Big] \Bigg\}\, .
\end{align}
The contribution due to the improvement term to the fermionic part of the
energy-momentum tensor is made up of 3 terms 
\be
{\cal T}^{F\{6\}(1,N_f)}_{\mu\nu} \longrightarrow {\cal T}^{F\{6\}(1,N_f)}_{\mu\nu} +
(1-\delta_{\mu\nu})\Big\{{\cal T}^{F4}_{\mu\nu} + {\cal T}^{F5}_{\mu\nu} + {\cal T}^{F6}_{\mu\nu}
\Big\}\; ,  
\ee
where
\bea
\langle T_{\mu\nu}^{F(0)} (S^{SW,1} S^{F,1}) \rangle_{\rm con} & = & (N_c^2-1) N_f\, {\cal T}^{F4}_{\mu\nu}\; ,\\
\langle T_{\mu\nu}^{F(0)} {(S^{SW,1})^2\over2} \rangle_{\rm con} & = & (N_c^2-1) N_f\, {\cal T}^{F5}_{\mu\nu}\; , \\
\langle T_{\mu\nu}^{F(1)} (-S^{SW,1}) \rangle & = & (N_c^2-1) N_f\, {\cal T}^{F6}_{\mu\nu}\; .
\eea
The expressions of ${\cal T}^{F4}_{\mu\nu}$,${\cal T}^{F5}_{\mu\nu}$, and ${\cal T}^{F6}_{\mu\nu}$ in terms of
integrals defined in appendix \ref{integrals} are
\begin{align}
& {\cal T}^{F4}_{\mu\nu} = - 2 a c_{\rm sw} \int_{q_{_{\vec\xi}};p_{_{\vec\xi,\theta}};k_{_{\vec\xi,\theta}}}\!\!\! 
{\bar\delta(p-q-k)\over D_G(q)D^2_F(k)D_F(p)}\Bigg\{ \bar k_\mu \bar k_\nu\bigg[
a \sum_{\rho\sigma} \Big( \big(\bar p_\rho + \bar k_\rho \big) \bar q_\sigma (\bar p_\rho  \bar k_\sigma -
\bar p_\sigma \bar k_\rho) \Big)
\nonumber \\
& + \sum_{\rho\sigma \atop \rho\neq\sigma} \bar q_\sigma \Big(m_0(k) \bar p_\sigma - m_0(p) \bar k_\sigma \Big)
\Big( \crh(p) + \crh(k)\Big)\bigg] +
{D_F(k) \over 4}\bigg\{\bar k_\nu\bigg[
  a \Big( \bar p_\mu + \bar k_\mu \Big) \sum_{\sigma} \bar p_\sigma \bar q_\sigma\nonumber\\
& -a \bar q_\mu \sum_{\sigma} \Big( \bar p_\sigma + \bar k_\sigma \Big) \bar p_\sigma
 +m_0(p) \bar q_\mu \sum_{\sigma\neq\mu} \Big( \cs(p) + \cs(k)\Big)
\bigg] + (\mu\leftrightarrow\nu)
\bigg\}\Bigg\}\; , 
\end{align}
\begin{align}
& {\cal T}^{F5}_{\mu\nu} =  a^2 c_{\rm sw}^2 \int_{q_{_{\vec\xi}};p_{_{\vec\xi,\theta}};k_{_{\vec\xi,\theta}}}\!\!\! {\bar\delta(p-q-k)
\over D_G(q) D_F^2(k) D_F(p)}\Bigg\{\bar k_\mu \bar k_\nu \bigg\{
{1\over 2} \Big[ \sum_\sigma \bar p_\sigma \bar k_\sigma - m_0(k) m_0(p) \Big]\times
\nonumber \\ 
&
\Big[\sum_\sigma \bar q^2_\sigma \Big( 3 +\sum_{\lambda\neq\sigma} \cl (q) \Big)\Big]
- \sum_\sigma \bar p_\sigma \bar q_\sigma  \sum_\rho \bar k_\rho  \bar q_\rho
\Big(2 - \cs (q)  + \sum_{\lambda\neq\rho} \cl (q)\Big)\nonumber \\
& -\sum_\sigma \bar q^2_\sigma \sum_\lambda \bar k_\lambda \bar p_\lambda \Big(1+ \cl (q) \Big) \bigg\}
- {D_F(k) \over 4}\bigg\{\bar k_\nu\bigg[
{1\over 2} \bar p_\mu \sum_\sigma \Big[ \bar q^2_\sigma \Big(1 - 2 \cm (q) + \sum_{\lambda\neq\sigma} \cl (q)\Big)\Big]\nonumber\\
&- \bar q_\mu \Big[ \sum_\sigma \bar p_\sigma \bar q_\sigma \Big( 2-\cm(q) + \sum_{\lambda\neq\sigma} \cl (q) \Big)\Big]\bigg]
+(\mu \leftrightarrow \nu)\bigg\}\Bigg\}\; ,  
\end{align}
and
\begin{align}
& {\cal T}^{F6}_{\mu\nu} =\!
{a c_{\rm sw} \over4}\!\! \int_{q_{_{\vec\xi}};p_{_{\vec\xi,\theta}};k_{_{\vec\xi,\theta}}}\!\!\!\! 
{\bar\delta(p-q-k)\over D_G(q)D_F(k)D_F(p)}
\Bigg\{\bigg\{\bar q_\mu \Big(\cn(p)\! +\! \cn (k) \Big) \Big[m_0(p)\bar k_\nu\! -\! m_0(k) \bar p_\nu\Big] \nonumber \\ %
& +(\mu \leftrightarrow \nu) \bigg\}
-2\delta_{\mu\nu} \Big(\cm(p)\! +\! \cm (k) \Big)
\sum_\sigma \Big[\bar q_\sigma (m_0(p) \bar k_\sigma\! -\! m_0(k) \bar p_\sigma)\Big]\Bigg\}\; . 
\end{align}

\section{Expectation values of triplet component of $T_{\mu\nu}$ \label{app:freeT3}}
Here we report the coefficients of the 1-loop perturbative expansion
of the expectation values of the triplet components of the energy-momentum tensor
defined in Eqs.~(\ref{eq:Tsextet})--(\ref{eq:T1lsextet}). The tree-level value of the
gluonic part is given by (no summation over repeated indices in this appendix unless
explicitly indicated)
\be
{\cal T}^{G\{3\}(0)}_{\mu\nu} =  \sum_{\alpha\neq \mu,\nu}
\Big[B^{(2)}_{\alpha\alpha\mu} - B^{(2)}_{\alpha\alpha\nu} + B^{(2)}_{\mu\mu\alpha} -  B^{(2)}_{\nu\nu\alpha} \Big]\; .   
\ee
The 1-loop contributions to the gluonic component are 
\bea
{\cal T}^{G\{3\}(1,N_c)}_{\mu\nu} & = & \Big\{ {\cal T}^{G1}_{\mu\mu} +
{\cal T}^{G2}_{\mu\mu} + {\cal T}^{G3}_{\mu\mu}\Big\} - (\mu\rightarrow\nu) \; ,  \\
{\cal T}^{G\{3\}(1,{1 \over N_c})}_{\mu\nu} & = & \Big\{
{\cal T}^{G4}_{\mu\mu} + {\cal T}^{G5}_{\mu\mu}\Big\} - (\mu\rightarrow\nu)  \; , \\
{\cal T}^{G\{3\}(1,N_f)}_{\mu\nu} & = & \Big\{
{\cal T}^{G6}_{\mu\mu} + {\cal T}^{G7}_{\mu\mu}\Big\} - (\mu\rightarrow\nu)\; . 
\eea
The tree-level expectation value of $T^{F, \{3\}}_{\mu\nu}$ is 
\begin{equation}
{\cal T}^{F\{3\}(0)}_{\mu\nu} = - 4 \big(F_{\mu\mu}^{(0)} - F_{\nu\nu}^{(0)}) \; , 
\end{equation}
and its derivative with respect to the bare mass is
\be\label{eq:DT3}
{\partial {\cal T}^{F\{3\}(0)}_{\mu\nu}  \over \partial m_0} =  8(F^{(1)}_{\mu\mu} - F^{(1)}_{\nu\nu})\; . 
\ee
The 1-loop fermion contribution is
\be
{\cal T}^{F\{3\}(1,N_f)}_{\mu\nu} =
\Big\{{\cal T}^{F1}_{\mu\mu} + {\cal T}^{F2}_{\mu\mu} + {\cal T}^{F3}_{\mu\mu}\Big\} - (\mu\rightarrow\nu)\; .
\ee

\subsection{O($a$) improvement}
The 1-loop contributions to the gluonic and fermionic parts due to the improvement are
\be
 {\cal T}^{G\{3\}(1,N_f)}_{\mu\nu} \longrightarrow {\cal T}^{G\{3\}(1,N_f)}_{\mu\nu} +
\Big\{{\cal T}^{G8}_{\mu\mu} + {\cal T}^{G9}_{\mu\mu} - (\mu\rightarrow\nu) \Big\}   
\ee   
and 
\be
{\cal T}^{F\{3\}(1,N_f)}_{\mu\nu} \longrightarrow {\cal T}^{F\{3\}(1,N_f)}_{\mu\nu} +
\Big\{{\cal T}^{F4}_{\mu\mu} + {\cal T}^{F5}_{\mu\mu} + {\cal T}^{F6}_{\mu\mu}
- (\mu\rightarrow\nu) \Big\}
\ee
respectively.

\section{Integrals\label{integrals}}
In this appendix we report the definitions of the tree-level integrals
which appear in the appendices \ref{app:freeE}, \ref{app:freeV},
\ref{app:freeT6}, and \ref{app:freeT3}. The functions $\cm(p)$,
$\sm(p)$, $D_G(p)$, $D_F(p)$ and $\int_{p_{_{\vec\xi,\theta}}}$ are
given in appendix \ref{eq:appD}. Repeated indices are not summed over.
\vspace{0.25cm}

\begin{minipage}[t]{0.4\textwidth}
  \begin{center}
    Fermionic integrals
  \end{center}
\begin{flalign}
F^{(0)}_{\mu\nu}=\int_{p_{_{\vec\xi,\theta}}}  {\bar p_\mu \bar p_\nu \over D_F(p)} &&
\end{flalign}
\begin{flalign}
F^{(1)}_{\mu\nu}=\int_{p_{_{\vec\xi,\theta}}}  {m_0(p) \bar p_\mu \bar p_\nu \over D_F^2(p)} &&
\end{flalign}
\end{minipage}
\hspace{0.75cm}\begin{minipage}[t]{0.35\textwidth}
  \begin{center}
    Bosonic integrals
  \end{center}
\begin{flalign}
  B^{(0)}=\int_{p_{_{\vec\xi}}}  {1\over D_G(p)}&&
\end{flalign}
\begin{flalign}
  B^{(1)}_{\mu\nu\alpha}=\int_{p_{_{\vec\xi}}}  {\bar p_\mu \bar p_\nu \ca^2 ({p\over 2})\over D_G^2(p)}&&
\end{flalign}
\end{minipage}

\begin{minipage}[t]{0.4\textwidth}
\vspace*{0.20cm}
\begin{flalign}
F^{(2)}_{\mu}= F^{(0)}_{\mu\mu} &&
\end{flalign}
\vspace*{0.05cm}
\begin{flalign}
F^{(3)}_{\mu}=\int_{p_{_{\vec\xi,\theta}}}  {m_0(p) \cm(p) \over D_F(p)} &&
\end{flalign}
\begin{flalign}
F^{(4)}_{\mu\nu}=\int_{p_{_{\vec\xi,\theta}}}  {\bar p_\mu \cn(p)\over D_F(p)} &&
\end{flalign}
\begin{flalign}
F^{(5)}_{\mu}=\int_{p_{_{\vec\xi,\theta}}}  {m_0(p) \bar p_\mu \over D_F(p)} &&
\end{flalign}
\begin{flalign}
 F^{(6)}_{\mu\nu}=\int_{p_{_{\vec\xi,\theta}}}  {m_0(p) \bar p_\mu \cn(p)\over D_F^2(p)} &&
\end{flalign}
 \begin{flalign}
 F^{(7)}_{\mu}=\int_{p_{_{\vec\xi,\theta}}}  {m_0^2(p) \bar p_\mu \over D_F^2(p)} &&
\end{flalign}
\begin{flalign}
F^{(8)} =\int_{p_{_{\vec\xi,\theta}}}  {m_0(p) \over D_F(p)} &&
\end{flalign}
\end{minipage}
\hskip 1cm 
\begin{minipage}[t]{0.45\textwidth}
\begin{flalign}
 B^{(2)}_{\mu\nu\alpha}=\int_{p_{_{\vec\xi}}}  {\bar p_\mu \bar p_\nu \ca^2 ({p\over 2})\over D_G(p)}&&
\end{flalign}
\begin{flalign}
  B^{(3)}_\mu=\int_{p_{_{\vec\xi}}}  {\cm(p)\over D_G(p)}&&
\end{flalign}
\begin{flalign}
B^{(4)}_{\mu\nu}=\int_{p_{_{\vec\xi}}}  {\bar p_\mu \bar p_\nu \over D_G(p)}&&
\end{flalign}
\begin{flalign}
  B^{(5)}_{\mu\nu\alpha\sigma}=\int_{p_{_{\vec\xi}}}  {\bar p_\mu \bar p_\nu \ca^2 ({p\over 2})\cs (p)\over D_G^2(p)}&&
\end{flalign}
\begin{flalign}
  B^{(6)}_{\mu\nu}=\int_{p_{_{\vec\xi}}}  {\cm(p) \cn(p) \over D_G(p)}&&
\end{flalign}
\end{minipage}





\bibliographystyle{JHEP}
\bibliography{bibfile.bib}

\end{document}